\documentclass[letterpaper]{JHEP3}

\usepackage{amsmath,amsthm, amssymb}

\newcommand{\dst}{\displaystyle}

\newcommand{\be}{\begin{equation}}
\newcommand{\ee}{\end{equation}}
\newcommand{\ba}{\begin{array}}
\newcommand{\ea}{\end{array}}

\newcommand{\bea}{\begin{eqnarray}}
\newcommand{\eea}{\end{eqnarray}}
\newcommand{\bma}{\begin{matrix}}
\newcommand{\ema}{\end{matrix}}
\newcommand{\bpm}{\begin{pmatrix}}
\newcommand{\epm}{\end{pmatrix}}
\newcommand{\nn}{\nonumber}
\newcommand{\half}{\frac{1}{2}}
\newcommand{\qter}{\frac{1}{4}}
\newcommand{\mc}{\mathcal}

\newcommand{\p}{\partial}

\newcommand{\rr}{\prime}
\newcommand{\ov}{\overline}
\newcommand{\wh}{\widehat}
\newcommand{\wt}{\widetilde}

\newcommand{\psibar}{\ov \psi}
\newcommand{\etabar}{\ov \eta}

\newcommand{\zetabar}{\ov \zeta}
\newcommand{\phibar}{\ov\phi}

\newcommand{\eps}{\varepsilon}
\newcommand{\ep}{\epsilon}

\newcommand{\al}{\alpha}

\newcommand{\la}{\lambda}
\newcommand{\da}{\delta}
\newcommand{\om}{\omega}

\newcommand{\ga}{\gamma}

\newcommand{\ta}{\omega}

\newcommand{\qrq}{\quad\Rightarrow\quad}

\newcommand{\epbar}{\ov\ep}

\newcommand{\slD}{{\hspace{2pt}\slash\hspace{-8pt}D}} 
\newcommand{\lra}{\leftrightarrow}

\title{\bf Simple $\bf d=4$ supergravity with a boundary}
\author{ Dmitry V.~Belyaev$^{a,b}$ and Peter van Nieuwenhuizen$^a$\\

$^a$ C.~N.~Yang Institute for Theoretical Physics, SUNY at Stony Brook \\
Stony Brook, NY 11794-3840, USA\\
\email{belyaev,vannieu@insti.physics.sunysb.edu}\\

$^b$ Deutsches Elektronen-Synchrotron, DESY-Theory\\ 
Notkestrasse 85, 22603 Hamburg, Germany\\
\email{dmitry.belyaev@desy.de}
}

\abstract{
To construct rigidly or locally supersymmetric bulk-plus-boundary actions, one needs an extension of the usual tensor calculus. Its key ingredients are the extended ($F$-, $D$-, etc.) density formulas and the rule for the decomposition of bulk multiplets into (co-dimension one) boundary multiplets. Working out these ingredients for $d=4$ $N=1$ Poincar\'e supergravity, we discover the special role played by $R$-symmetry (absent in the $d=3$ $N=1$ case we studied previously). The $U(1)_A$ $R$-symmetry has to be gauged which leads us to extend the old-minimal set of auxiliary fields $S,P,A_\mu$ by a $U(1)_A$ compensator $a$. Our results include the ``$F+A$'' density formula, the ``$Q+L+A$'' formula for the induced supersymmetry transformations (closing into the standard $d=3$ $N=1$ algebra) and demonstration that the compensator $a$ is the first component of the extrinsic curvature multiplet. We rely on the superconformal approach which allows us to perform, in parallel, the same analysis for new-minimal supergravity. 
}


\begin{document}

\numberwithin{equation}{section}


\section{Introduction} 

Rigid and local supersymmetry (susy) in the presence of boundaries have been studied before (see references in \cite{bvn1}), but in most of those studies boundary conditions (BC) were imposed in order that the action remains supersymmetric in the presence of a boundary. Often these BC were treated on a par with the BC one gets from the Euler-Lagrange variational equations. We are, instead, interested in constructing actions that remain susy without imposing any BC (``susy without BC'') by adding suitable boundary terms to the action \cite{bvn1,bvn2}.
For rigid susy, the formalism of co-dimension one superfields \cite{rigid54} provides an easy way to construct bulk-plus-boundary actions that are ``susy without BC'' \cite{db13,bvn2}. In \cite{bvn1} we showed that the same could be achieved in the local susy case by developing a boundary-extended tensor calculus for $d=3$ $N=1$ Poincar\'e supergravity (sugra). We found there that the standard $d=3$ $F$-density formula\footnote{
Our conventions are summarized in Appendix \ref{appC}. To avoid confusion with the auxiliary field $S$ in $d=4$, we denote the auxiliary field in $d=3$ by $S_3$.}
\bea
\mc{L}_F=e_3\Big( F+\half\psibar_M\ga^M\chi+\qter A\psibar_M\ga^{M N}\psi_N+A S_3 \Big)
\eea
for a scalar multiplet $(A,\chi,F)$ interacting with the supergravity multiplet $(e_M{}^A,\psi_M,S_3)$, combined with a boundary ``$A$-term,'' gives rise to a bulk-plus-boundary action (that we called the ``$F+A$'' formula)
\bea
\label{FplusA1}
S=\int_\mc{M} d^3x \mc{L}_F
-\int_{\p\mc{M}} d^2x e_2 A
\eea
which is supersymmetric under half of bulk susy parametrized by $\ep_{+}(x)=\half(1+\ga^{\hat 3})\ep(x)$, provided we use \emph{modified} susy transformations
\bea
\label{3dmsusy}
\da^\rr(\ep_{+})=\da_Q^P(\ep_{+})+\da_L(\la_{a\hat 3}=-\epbar_{+}\psi_{a-})
\eea
which close into the standard $d=2$ $N=(1,0)$ superalgebra. The accompanying (local) Lorentz transformation $\da_L(\la_{a\hat 3})$ we found to be needed as a compensating transformation to maintain a particular Lorentz gauge,
\bea
\label{egauge}
e_m{}^{\hat 3}=e_a{}^3=0
\eea
This gauge is opposite to the standard Kaluza-Klein gauge choice $e_3{}^a=e_{\hat 3}{}^m=0$. However, as we demonstrate explicitly in Appendix \ref{appA}, it is not necessary to impose this Lorentz gauge. By using projection operators familiar from the Gauss-Codazzi equations \cite{wald}, we define projected induced symmetry transformations that lead to the same results.

We will demonstrate that the program of ``susy without BC'' works for $d=4$ $N=1$ supergravity, provided we use a formulation which maintains the $U(1)_A$ (with ``$A$'' for ``axial'') $R$-symmetry as a \emph{local} symmetry. New-minimal sugra \cite{new-min} inherits the $U(1)_A$ gauge symmetry from conformal sugra \cite{KakuNT}, but in old-minimal sugra \cite{old-min} the $U(1)_A$ local symmetry has been gauge fixed. Thus we relax the $U(1)_A$ gauge, after which both old- and new-minimal sugra can be treated on a par. We will find, in particular, that the modified susy that is preserved by the boundary and which closes into the standard $d=3$ $N=1$ superalgebra has the form
\bea
\da^\rr(\ep_{+})=\da_Q^P(\ep_{+})
+\da_L(\la_{a\hat 3}(\ep_{+}))
+\da_A(\ta(\ep_{+}))
\eea
where the composite parameters $\la_{a\hat 3}(\ep_{+})$ and $\ta(\ep_{+})$ of the Lorentz and $U(1)_A$ transformation, respectively, depend on the bulk fields only through $\psi_{a-}$. This field $\psi_{a-}$ is supercovariant under local $\ep_{+}$ susy transformations.

\section{The ``$B$ problem''} 
\label{sec-Bprob}

When one tries to extend the $d=3$ $N=1$ results of \cite{bvn1} to the case of $d=4$ $N=1$ old-minimal (Poincar\'e) supergravity, one runs into the following problem. The $d=4$ $F$-density for a scalar (chiral) multiplet $(A,B,\chi,F^\rr,G^\rr)$ interacting with the supergravity multiplet $(e_M{}^A,\psi_M,S,P,A_M^\text{aux})$ reads \cite{tensor}
\bea
\label{PFD}
\mc{L}_F=e_4\Big[ F^\rr+\half\psibar_M\ga^M\chi
+\qter\psibar_M\ga^{M N}(A-i\ga_5 B)\psi_N+A S+B P \Big]
\eea
Under local susy ($\da e_M{}^A=\epbar\ga^A\psi_M$, $\da\psi_M=2\p_M\ep+\dots$, $\da A=\epbar\chi$, $\da B=-\epbar i\ga_5\chi$, etc.) it varies into a total derivative,
\bea
\label{totder}
\da(\ep)\mc{L}_F=\p_M\Big\{ e_4\Big[\epbar\ga^M\chi+\epbar\ga^{M N}(A-i\ga_5 B)\psi_N \Big] \Big\}
\eea
In the gauge (\ref{egauge}) and considering only the half of susy parametrized by $\ep_{+} \equiv \half(1+\ga^{\hat 3})\ep(x)$ with constant $\ga^{\hat 3}$, the variation of $\int_\mc{M} d^4x \mc{L}_F$ gives (we take $x^3\geq 0$ in $\mc{M}$ and the boundary $\p\mc{M}$ at $x^3=0$)
\bea
\label{bvar}
\int_{\p\mc{M}} d^3x e_3\Big[ \epbar_{+}\chi_{-}+\epbar_{+}\ga^a(\psi_{a+}A-i\ga_5\psi_{a-}B)\Big]
\eea
where we defined $\psi_a \equiv e_a{}^m\psi_m$, $\psi_{a\pm} \equiv \half(1\pm\ga^{\hat 3})\psi_a$ and used $\epbar_{+}\ga^{\hat 3}=-\epbar_{+}$. We also used that, in the gauge (\ref{egauge}), $e_3=e_{\hat 3}{}^3 e_4$ (with $e_3=\det e_m{}^a$ and $e_4=\det e_M{}^A$). From $\da e_3=e_3(\epbar_{+}\ga^a\psi_{a+})$ and $\da A=\epbar_{+}\chi_{-}$, one finds a natural candidate for the bulk-plus-boundary action
\bea
S_{F+A}=\int_\mc{M} d^4x \mc{L}_F
-\int_{\p\mc{M}} d^3x e_3 A
\eea
Its $\ep_{+}$ susy variation cancels the first two terms in (\ref{bvar}) but the $B$-term remains
\bea
\da(\ep_{+}) S_{F+A}=\int_{\p\mc{M}} d^3x e_3(\epbar_{+}i\ga_5\ga^a\psi_{a-}) B
\eea
This poses a problem as it appears to be impossible to cancel this remaining variation (without imposing any BC) within old-minimal supergravity.

A simple observation guides us towards the solution of this problem. Namely, the ``$F+A$'' action would be invariant if we could modify the $\ep_{+}$ susy transformation by an additional transformation rotating $A$ into $B$ with a composite parameter proportional to $\epbar_{+}i\ga_5\ga^a\psi_{a-}$. The bulk $F$-density should then be invariant under such a rotation. Recalling that old-minimal sugra is a gauge-fixed version of conformal supergravity where such a $U(1)_A$ symmetry has been gauge-fixed, we reconsider the gauge-fixing procedure and restore the $U(1)_A$ symmetry. One way of doing this would lead us to the known new-minimal formulation; another way produces a new version of old-minimal supergravity with an additional $U(1)_A$ compensator arising as a St\"uckelberg (or Goldstone) field.

\section{Poincar\'e sugras preserving $\bf U(1)_A$} 

All known $d=4$ $N=1$ Poincar\'e sugras follow from the superconformal approach by combining the gauge multiplet $(e_M{}^A,\psi_M,A_M,b_M)$ of conformal sugra \cite{KakuNT} with different conformal matter multiplets serving as ``compensator multiplets'' \cite{KT,PVN,KU}. Certain components of these multiplets are gauge fixed to break extraneous superconformal symmetries (such as dilatations) while the remaining components become auxiliary fields in the corresponding Poincar\'e sugra multiplet. Using a conformal chiral compensator multiplet $(A_0,B_0,\chi_0, F_0, G_0)$ one finds the old-minimal (OM) sugra multiplet $(e_M{}^A,\psi_M,A_M^\text{aux},S,P)$ with $A_M^\text{aux}=-\frac{3}{2}A_M$, $S=3F_0$ and $P=-3G_0$ \cite{KT}, while using a linear compensator multiplet $(C_0,\chi_0,\wt B_M)$ one finds new-minimal (NM) sugra with multiplet $(e_M{}^A,\psi_M,A_M,\wt B_M)$ \cite{KU}.\footnote{
The auxiliary vector $\wt B_M$ of NM sugra satisfies a constraint that can be explicitly solved in terms of a prepotential $a_{M N}$ which introduces an additional gauge symmetry in the algebra. For our discussion, the introduction of $a_{M N}$ is not necessary.
}
However, as we show in Appendix \ref{appB}, one can find further consistent (extended Poincar\'e) sugras by relaxing some of the gauge fixing conditions. 

Aimed at the problem at hand, we have relaxed the $U(1)_A$ gauge fixing condition imposed in the derivation of OM sugra and derived an ``old-minimal sugra with a $U(1)_A$ compensator'' (OMA sugra, for short). Its sugra multiplet is $(e_M{}^A,\psi_M,A_M,S,P,\phi)$, where $\phi$ is the $U(1)_A$ compensator. Under (Poincar\'e) susy transformations $\da(\ep)=\da_Q^P(\ep)$, Lorentz transformations $\da(\la)=\da_L(\la^{A B})$ and $U(1)_A$ transformations $\da(\ta)=\da_A(\ta)$ it varies as follows (see (\ref{secconstr}))
\bea
\label{singlet}
\da(\ep,\la,\ta) \phi=\ta
\eea
so that it is a susy singlet, a Lorentz scalar and a Goldstone boson of $U(1)_A$. As it can be gauge fixed to $\phi=0$ by a local $U(1)_A$ transformation, this is still a minimal (12+12) Poincar\'e sugra. 

We have thus two minimal sugras with local $U(1)_A$ symmetry: OMA and NM sugras. Their (Poincar\'e) susy transformations are encoded in the ``$Q+S+K$'' formula (\ref{QSK1})
\bea
\da_Q^P(\ep)=\da_Q(\ep)+\da_S(\zeta(\ep))+\da_K(\xi_K^A(\ep))
\eea
where the parameter in the $K$ (special conformal) transformation need not be specified as all the (independent) fields are inert under it. The key information is contained in the parameter $\zeta(\ep)$ of the $S$ (conformal susy) transformation. For OMA and NM sugras it is given by (see equation (\ref{OMAzeta1}) for OMA, and equation (3.28) in \cite{KU} for NM sugra)\footnote{
In OMA sugra we can replace $(A_M,\phi)$ by $(A_M^\text{aux},a)=-\frac{3}{2}(A_M,\phi)$; then $B_M=A_M^\text{aux}-\p_M a$.}
\bea
\label{zetas}
\text{OMA: } &&
\zeta(\ep)=H\ep, \quad
H=-\frac{1}{3}(S-i\ga_5 P+i\ga_5\ga^A B_A), \quad
B_M=-\frac{3}{2}(A_M-\p_M \phi) \nn\\
\text{NM: } &&
\zeta(\ep)=\wt H \ep, \quad
\wt H=\half i\ga_5 \ga^A \wt B_A, \quad
\wt B_M=\wt B_M(a_{M N})
\eea
We state explicitly only the $\da(\ep,\la,\ta)$ transformation rules of $e_M{}^A$ and $\psi_M$,
\bea
\label{epsitr}
\da(\ep,\la,\ta)e_M{}^A &=& \epbar\ga^A\psi_M+\la^{A B} e_{M B} \nn\\
\da(\ep,\la,\ta)\psi_M &=& 2 D(\wh\om)_M\ep-\frac{3}{2}i\ga_5\ep A_M-\ga_M\zeta(\ep)
+\qter\la^{A B}\ga_{A B}\psi_M+\frac{3}{4}i\ga_5\psi_M \ta 
\eea
which holds true in both OMA and NM sugras.

\section{Solution of the ``$B$ problem''} 

Both in OMA and NM sugras, unlike in the case of OM sugra, there is no need to redefine the fields $F,G$ of a chiral multiplet with $U(1)_A$ weight $n$ into fields $F^\rr,G^\rr$ whose transformation rules are $n$-independent, and in both cases we can use the same conformal $F$-density formula\footnote{
As $F^\rr=F-\frac{n}{3}(S A+P B)$, $G^\rr=G-\frac{n}{3}(S B-P A)$ \cite{tensor}, the Poincar\'e $F$-density (\ref{PFD}) coincides with the conformal $F$-density (\ref{CFD}) when $n=3$.}
\bea
\label{CFD}
\mc{L}_F=e_4\Big[ F+\half\psibar_A\ga^A\chi+\qter\psibar_A\ga^{A B}(A-i\ga_5 B)\psi_B \Big]
\eea
Under $U(1)_A$ the fields appearing in this density transform as
\bea
&&\da e_M{}^A=0, \quad
\da\psi_M=\frac{3}{4}i\ga_5\psi_M\ta, \quad
\da F=-\Big( \frac{n-3}{2} \Big)\ta G
\nn\\
&&\da A=-\frac{n}{2}\ta B, \quad
\da B=+\frac{n}{2}\ta A, \quad
\da\chi=i\Big( \frac{n}{2}-\frac{3}{4}\Big)\ga_5\chi \ta
\eea
and one can check that $\mc{L}_F$ is $U(1)_A$ invariant, $\da(\ta)\mc{L}_F=0$, provided $n=3$. Under (Poincar\'e) supersymmetry $\da(\ep)\equiv\da_Q^P(\ep)$ it transforms into a total derivative\footnote{
Note that $\da_Q^P(\ep)\mc{L}_F=\da_Q(\ep)\mc{L}_F$ as $\mc{L}_F$ is $K$-invariant for any $n$ (all fields in $\mc{L}_F$ are $K$-invariant) and $S$-invariant when $n=3$. 
(Under $S$-supersymmetry, $\da A=\da B=0$, $\da\chi=n(A+i\ga_5 B)\zeta$, $\da F=(1-n)\zetabar\chi$ and $\da e_M{}^A=0$, $\da\psi_M=-\ga_M\zeta$, $\da A_M=\zetabar i\ga_5\psi_M$, see \cite{PVN,KU}.) The action is also Weyl invariant for $n=3$.}
\bea
\da(\ep)\mc{L}_F=\p_M\Big\{ e_4\Big[\epbar\ga^M\chi+\epbar\ga^{M N}(A-i\ga_5 B)\psi_N \Big] \Big\}
\eea
which is of exactly the same form as in OM sugra, see (\ref{totder}). From here on the discussion of Section \ref{sec-Bprob} applies, but now we can resolve the problem encountered there by an extra ($\ep_{+}$)-dependent $U(1)_A$ transformation. For the ``$F+A$'' action in (\ref{FplusA1})
\bea
\label{FplusA}
S_{F+A}=\int_\mc{M} d^4x \mc{L}_F
-\int_{\p\mc{M}} d^3x e_3 A
\eea
the combined $\da(\ep_{+},\ta)\equiv\da_Q^P(\ep_{+})+\da_A(\om)$ variation gives
\bea
\da(\ep_{+},\ta) S_{F+A}=\int_{\p\mc{M}} d^3x e_3\Big[
(\epbar_{+}i\ga_5\ga^a\psi_{a-}) B+\frac{3}{2}\ta B \Big]
\eea
and it vanishes provided $\ta(\ep_{+})=-\frac{2}{3}(\epbar_{+}i\ga_5\ga^a\psi_{a-})$. This way the ``$B$ problem'' is solved.

\section{(Modified) induced susy transformations} 

Let us summarize what we have learned so far. First, in the presence of a boundary half of susy is (spontaneously) broken and for this reason we consider only $\da(\ep_{+})$ \cite{bvn1}. Second, in the gauge $e_m{}^{\hat 3}=e_a{}^3=0$ we need a compensating Lorentz transformation $\da(\la_{a\hat 3}=-\epbar_{+}\psi_{a-})$ \cite{bvn1}; another, gauge-independent reason for this modification is explained in Appendix \ref{appA}. Third, from the resolution of the ``$B$ problem,'' which arose in the construction of the supersymmetric bulk-plus-boundary ``$F+A$'' density formula, we found that we need a further modification, an additional $U(1)_A$ transformation $\da(\ta=-\frac{2}{3}\epbar_{+}i\ga_5\ga^a\psi_{a-})$. Putting all the pieces together, we claim that
\bea
\label{modsusy}
\da^\rr(\ep_{+})=\da_Q^P(\ep_{+})+\da_L\Big( \la_{a\hat 3}(\ep_{+})=-\epbar_{+}\psi_{a-} \Big)
+\da_A\Big( \ta(\ep_{+})=-\frac{2}{3}(\epbar_{+}i\ga_5\ga^a\psi_{a-}) \Big)
\eea
is the correct (modified) induced susy transformation which one should consider as surviving in the presence of the boundary. By ``correct'' we mean that this susy transformation closes into the standard $d=3$ $N=1$ susy algebra, as we now explicitly demonstrate.

From the superconformal algebra and the ``$Q+S+K$'' rule, we find that the commutator of two Poincar\'e susy transformations is given by (see (\ref{QPQP1}) with $\ep$ rescaled to $2\ep$)
\bea
[\da_Q^P(\ep_1),\da_Q^P(\ep_2)] &=& \da_{g.c.}(\xi^M)
+\da_Q^P(-\half\xi^A\psi_A)
+\da_L(\xi^C\wh\om_C{}^{A B}+\epbar_{[1}\ga^{A B}\zeta_{2]}) \nn\\
&& +\da_A(-\xi^A A_A+2\epbar_{[1}i\ga_5\zeta_{2]})
\eea
where $\xi^A=2\epbar_2\ga^A\ep_1$, $\xi^M=\xi^A e_A{}^M$, $\zeta_{1,2}=\zeta(\ep_{1,2})$ with $\zeta(\ep)$ given in (\ref{zetas}), and $[1 2]=1 2-2 1$. From the ``$Q+L+A$'' form of the (modified) induced susy transformation we further find
\bea
\label{1pr2pr}
[\da^\rr(\ep_{1+}),\da^\rr(\ep_{2+})] &=& \da_{g.c.}(\xi^M)
+\da_Q^P(\ep_3)+\da_L(\la_3{}^{A B})+\da_A(\ta_3) \nn\\
&&
\ep_{3}= -\half\xi^A\psi_A
+\qter\ga_{A B}\ep_{[1+}\la_{2]}{}^{A B}+\frac{3}{4}i\ga_5\ep_{[1+}\ta_{2]} \nn\\[5pt]
&&
\la_3{}^{A B}= \xi^C\wh\om_C{}^{A B}+\epbar_{[1+}\ga^{A B}\zeta_{2]}
+\la_{[2}{}^{A C}\la_{1] C}{}^B+\da^\rr(\ep_{[1+})\la_{2]}{}^{A B} \nn\\[5pt]
&& 
\ta_3=-\xi^A A_A+2\epbar_{[1}i\ga_5\zeta_{2]}
+\da^\rr(\ep_{[1+})\ta_{2]} 
\eea
where we took into account the field-dependence in composite parameters of Lorentz and $U(1)_A$ transformations and denoted $\zeta_2\equiv\zeta(\ep_{2+})$, etc. Using $\la^{a b}(\ep_{+})=0$ and the form of $\la^{a\hat 3}(\ep_{+})$ and $\ta(\ep_{+})$ in (\ref{modsusy}), we find for the composite parameters
\bea
\label{composites}
&& \xi^a=2(\epbar_{1+}\ga^a\ep_{2+}), \quad
\xi^{\hat 3}=2(\epbar_{1+}\ga^{\hat 3}\ep_{2+})=0 \qrq
\xi^m=\xi^a e_a{}^m, \quad
\xi^3=0 \nn\\
&& \ep_3=-\half\xi^a\psi_a+\wt\ep, \quad
\la_3{}^{A B}=\xi^c\wh\om_c{}^{A B}+\wt\la^{A B}, \quad
\ta_3=-\xi^a A_a+\wt\ta
\eea
where we separated parts that need more work,
\bea
\wt\ep &\equiv& \half\ga^{a\hat 3}\ep_{1+}\la_{a\hat 3}(\ep_{2+})
+\frac{3}{4}i\ga_5\ep_{1+}\ta(\ep_{2+}) -(1\lra 2) 
\nn\\
\wt\la_{a b} &\equiv& -\epbar_{2+}\ga_{a b}\zeta_{-}(\ep_{1+})
-\la_{a\hat 3}(\ep_{2+})\la_{b\hat 3}(\ep_{1+})-(1\lra 2) 
\nn\\[5pt]
\wt\la_{a\hat 3} &\equiv& -\epbar_{2+}\ga^{a\hat 3}\zeta_{+}(\ep_{1+})
-\epbar_{2+}\da^\rr(\ep_{+})\psi_{a-}-(1\lra 2) 
\nn\\
\wt\ta &\equiv& -2\epbar_{2+}i\ga_5\zeta(\ep_{1+})
-\frac{2}{3}\epbar_{2+}i\ga_5\ga^a\da^\rr(\ep_{+})\psi_{a-}-(1\lra 2)
\eea
After some Fierzing, we find (see Appendix \ref{appC})
\bea
\label{final-ep3}
\wt\ep=\half\xi^a\psi_{a-} \qrq \ep_3=-\half\xi^a\psi_{a+} \qrq
\ep_{3+}=-\half\xi^a\psi_{a+}, \quad
\ep_{3-}=0
\eea
Writing $\wh\om_{m a b}=\wh\om_{m a b}^{+}+\kappa_{m a b}^{-}$ (where $\wh\om_{m a b}^{+}$ depends only on $\psi_{m+}$ and $\kappa_{m a b}^{-}$ is the part of the contorsion that depends only on $\psi_{m-}$), Fierzing and using that the complete antisymmetrization of three $d=3$ indices must be proportional to the $d=3$ Levi-Civita tensor, we find (see Appendix \ref{appC})
\bea
\label{la3ab}
\la_3{}^{a b}=\xi^c\wh\om_c^{+}{}^{a b}-\epbar_{[2+}\ga^{a b}\zeta^\rr_{-}(\ep_{1]+}), \quad
\zeta^\rr(\ep_{+})\equiv\zeta(\ep_{+})+\frac{1}{8}i\ga_5\ep_{+}(\psibar_{a-}i\ga_5\ga^{a b}\psi_{b-})
\eea
In the next section, we will further simplify this expression using the explicit form of $\zeta(\ep)$ for OMA and NM sugras.

To work out $\la_3{}^{a\hat 3}$ and $\ta_3$, we need first to determine $\da^\rr(\ep_{+})\psi_{a-}$.
As a warm up exercise, we evaluate $\da^\rr(\ep_{+})e_m{}^a$ and $\da^\rr(\ep_{+})e_a{}^m$. We have
\bea
\da^\rr(\ep_{+})e_m{}^a &=& \epbar_{+}\ga^a\psi_{m}+\la^{a\hat 3}e_m{}^{\hat 3}
=\epbar_{+}\ga^a\psi_{m+} \nn\\
\da^\rr(\ep_{+})e_a{}^m &=& -\epbar_{+}\ga^m\psi_{a}+\la_{a\hat 3}e_{\hat 3}{}^m \nn\\
&=& -\epbar_{+}(\ga^b e_b{}^m+\ga^{\hat 3}e_{\hat 3}{}^m)\psi_{a}-(\epbar_{+}\psi_{a-})e_{\hat 3}{}^m 
=-\epbar_{+}\ga^b\psi_{a+} e_b{}^m
\eea
where we used $e_m{}^{\hat 3}=0$, which is our gauge choice (\ref{egauge}); but note that $e_{\hat 3}{}^m \neq 0$. For $\da^\rr(\ep_{+})$ of $\psi_{a-}=e_a{}^m P_{-}\psi_m$, with $P_{-}=\half(1-\ga^{\hat 3})$, we have, using (\ref{epsitr}),
\bea
\label{dapsiam1}
\da^\rr(\ep_{+})\psi_{a-} &=& \Big[\da^\rr(\ep_{+}) e_a{}^m \Big] \psi_{m-}
+e_a{}^m P_{-}\Big[
2\Big( \p_m+\frac{1}{4}\wh\om_{m c b}\ga^{c b}+\half\wh\om_{m c\hat 3}\ga^{c\hat 3} \Big)\ep_{+}
\nn\\
&& -\frac{3}{2}i\ga_5\ep_{+}A_m-\ga_m\zeta(\ep_{+})
+\half\ga_{c\hat 3}\psi_m\la^{c\hat 3}(\ep_{+})+\frac{3}{4}i\ga_5\psi_m\ta(\ep_{+}) \Big]
\eea
As $P_{-}\ep_{+}=0$, the term with $\p_m\ep_{+}$ is projected out. This shows that $\psi_{a-}$ is supercovariant under $\da^\rr(\ep_{+})$ susy. We expect supercovariant quantities to transform into supercovariant quantities, and we find that this is indeed the case (see Appendix \ref{appC}):
\bea
\label{dapsiam}
\da^\rr(\ep_{+})\psi_{a-}=\ga^b\ep_{+}\wh K_{a b}-\frac{3}{2}i\ga_5\ep_{+}\wh A_a-\ga_a\zeta_{+}(\ep_{+})
\eea
where $\wh K_{a b}$ is the supercovariant extrinsic curvature tensor and $\wh A_a$ is the supercovariant $d=3$ vector part of the $U(1)_A$ gauge vector $A_A$, 
\bea
\label{scKA}
\wh K_{a b}=\wh\om_{a b\hat 3}-\half\psibar_{a+}\psi_{b-}, \quad
\wh A_a=A_a+\frac{1}{3}\psibar_{a+}i\ga_5\ga^b\psi_{b-}
\eea
We emphasize that the supercovariance is with respect to the (modified) induced susy transformation $\da^\rr(\ep_{+})$; for example, 
\bea
\da^\rr(\ep_{+})A_a=\da_A(\ta(\ep_{+}))A_a+\dots
=\p_a\ta(\ep_{+})+\dots
=-\frac{2}{3}(\p_a\epbar_{+})i\ga_5\ga^b\psi_{b-}+\dots 
\nn
\eea
leads us to $\wh A_a$ (as $\da\psi_{a+}=2\p_a\ep_{+}+\dots$). 

For $\wt\la_{a\hat 3}$ we now have
\bea
\wt\la_{a\hat 3}=-\epbar_{2+}\ga^a\zeta_{+}(\ep_{1+})
-\epbar_{2+}\Big[
\ga^b\ep_{1+}\wh K_{a b}-\frac{3}{2}i\ga_5\ep_{1+}\wh A_a-\ga_a\zeta_{+}(\ep_{1+}) \Big]
-(1\lra 2)
\eea
We see that the two $\zeta$-dependent terms cancel while $\wh A_a$-dependent term vanishes due to ``$1\lra 2$.'' Using $\xi^a=2(\epbar_{2+}\ga^a\ep_{1+})$ we find $\wt\la_{a\hat 3}=-\xi^b\wh K_{a b}$, while using $\wh K_{a b}=\wh K_{b a}$ we obtain the final result
\bea
(\la_3)_{a\hat 3}=-\xi^b(\wh K_{b a}-\wh\om_{b a\hat 3})
=\half\xi^b\psibar_{b+}\psi_{a-}=\la_{a\hat 3}(\ep_{3+})
\eea
where $\la_{a\hat 3}(\ep_{+})=-\epbar_{+}\psi_{a-}$ and $\ep_{3+}=-\half\xi^b\psi_{b+}$ according to (\ref{final-ep3}).

The calculation of $\ta_3$ is equally simple. We start with
\bea
\wt\ta=-2\epbar_{2+}i\ga_5\zeta_{+}(\ep_{1+})-\frac{2}{3}\epbar_{2+}i\ga_5\ga^a\Big[
\ga^b\ep_{1+}\wh K_{a b}-\frac{3}{2}i\ga_5\ep_{1+}\wh A_a-\ga_a\zeta_{+}(\ep_{1+}) \Big]
-(1\lra 2)
\eea
Again, the $\zeta$-dependent terms cancel, now thanks to $\ga^a\ga_a=3$. The term with $\wh K_{a b}$ vanishes due to $\wh K_{a b}=\wh K_{b a}$ and ``$1\lra 2$.'' Using $\ga_5\ga^a\ga_5=-\ga^a$, we find $\wt\ta=\xi^a\wh A_a$. This finally gives
\bea
\ta_3=\xi^a(\wh A_a-A_a)=\frac{1}{3}\xi^a(\psibar_{a+}i\ga_5\ga^b\psi_{b-})=\ta(\ep_{3+})
\eea
where $\ta(\ep_{+})=-\frac{2}{3}\epbar_{+}i\ga^5\ga^b\psi_{b-}$ and $\ep_{3+}=-\half\xi^a\psi_{a+}$.

We now collect our findings. For the commutator (\ref{1pr2pr}) of two (modified) induced susy transformations, we obtain
\bea
[\da^\rr(\ep_{1+}),\da^\rr(\ep_{2+})]=\da_{g.c.}(\xi^a e_a{}^m)+\da_Q^P(\ep_{3+})
+\da_L(\la_3^{a b})+\da_L(\la_{a\hat 3}(\ep_{3+}))+\da_A(\ta(\ep_{3+}))
\eea
where $\xi^a=2(\epbar_{2+}\ga^a\ep_{1+})$, $\ep_{3+}=-\half\xi^a\psi_{a+}$ and $\la_3{}^{a b}$ is given in (\ref{la3ab}). We observe that the (unmodified) Poincar\'e susy, off-diagonal Lorentz and the $U(1)_A$ transformations on the right hand side recombine into the (modified) induced susy transformation $\da^\rr(\ep_{3+})$ and the result is simply
\bea
[\da^\rr(\ep_{1+}),\da^\rr(\ep_{2+})]=\da_{g.c.}(\xi^a e_a{}^m)
+\da^\rr(-\half\xi^a\psi_{a+})
+\da_L(\la_3^{a b})
\eea
Up to some final simplification of $\la_3^{a b}$ and decomposition of 4-component spinors $\ep_{+}$ and $4\times 4$ gamma matrices $\ga^a$ in terms of 2-component spinors and $2\times 2$ gamma matrices, which will be done in the next section, this is the correct $d=3$ $N=1$ susy algebra confirming our claim that $\da^\rr(\ep_{+})$ in (\ref{modsusy}) is the correct expression for the induced susy transformations.

\section{Induced sugra multiplets in OMA and NM sugra} 

Our ``$F+A$'' action formula (\ref{FplusA}) gives one possible $\da^\rr(\ep_{+})$-supersymmetric bulk-plus-boundary completion of the bulk $F$-density formula. However, other possibilities exist because we can add further, separately $\da^\rr(\ep_{+})$-supersymmetric, boundary actions. To construct such boundary actions and to obtain an explicit boundary $F$-density formula, we need first to find the induced sugra multiplet.

We have found already that $\da^\rr(\ep_{+})e_m{}^a=\epbar_{+}\ga^a\psi_{m+}$. To identify the combination of bulk fields which plays the role of the $d=3$ auxiliary field $S_3$ in the induced sugra multiplet $(e_m{}^a,\psi_{m+},S_3)$ we need to work out $\da^\rr(\ep_{+})\psi_{m+}$. Using $\psi_{m+}=P_{+}\psi_m$, $P_{+}=\half(1+\ga^{\hat 3})$ and (\ref{epsitr}), we write
\bea
\label{dapsiap1}
\da^\rr(\ep_{+})\psi_{m+} &=& P_{+}\Big[
2\Big( \p_m+\frac{1}{4}\wh\om_{m a b}\ga^{a b}+\half\wh\om_{m a\hat 3}\ga^{a\hat 3} \Big)\ep_{+}
\nn\\
&& -\frac{3}{2}i\ga_5\ep_{+}A_m-\ga_m\zeta(\ep_{+})
+\half\ga_{a\hat 3}\psi_m\la^{a\hat 3}(\ep_{+})+\frac{3}{4}i\ga_5\psi_m\ta(\ep_{+}) \Big]
\eea
Now it is the $\wh\om_{m a\hat 3}$ and $A_m$ dependent terms which are projected out. For the remaining terms, after some algebra very similar to that used in deriving (\ref{la3ab}), we find (see Appendix~\ref{appC})
\bea
\label{dapsiap}
\da^\rr(\ep_{+})\psi_{m+}=2D^\rr(\wh\om^{+})_m \ep_{+}-\ga_m\zeta_{-}^\rr(\ep_{+}), \quad
\zeta^\rr(\ep_{+})\equiv\zeta(\ep_{+})+\frac{1}{8}i\ga_5\ep_{+}(\psibar_{a-}i\ga_5\ga^{a b}\psi_{b-})
\eea
where $D^\rr(\wh\om^{+})_m$ is the induced supercovariant derivative,
\bea
D^\rr(\wh\om^{+})_m=\p_m+\qter\wh\om_{m a b}^{+}\ga^{a b}, \quad
\wh\om_{m a b}^{+}=\om(e)_{m a b}+\kappa_{m a b}^{+}
\eea
Using the explicit form of $\zeta(\ep)$ for OMA and NM sugras, see (\ref{zetas}), we find
\bea
\label{zetaspm}
\text{OMA: } &&
\zeta_{+}(\ep_{+})=-\frac{1}{3}(S+i\ga_5\ga^a B_a)\ep_{+}, \quad
\zeta_{-}(\ep_{+})=\frac{1}{3}i\ga_5\ep_{+}(P-B_{\hat 3}) \nn\\
\text{NM: } &&
\zeta_{+}(\ep_{+})=\half i\ga_5\ga^a\ep_{+} \wt B_a, \quad
\zeta_{-}(\ep_{+})=\half i\ga_5\ep_{+} \wt B_{\hat 3}
\eea
Therefore, in both cases we can write $\zeta_{-}^\rr(\ep_{+})=-\half i\ga_5 \ep_{+} S_3$ so that
\bea
\da^\rr(\ep_{+})\psi_{m+}=2D^\rr(\wh\om^{+})_m\ep_{+}-\half i\ga_5\ga_m\ep_{+} S_3
\eea
where $S_3$ is given by
\bea
\text{OMA: } && S_3=-\frac{2}{3}(P-B_{\hat 3})-W, \quad
W\equiv \qter\psibar_{a-}i\ga_5\ga^{a b}\psi_{b-} \nn\\
\text{NM: } && S_3=-\wt B_{\hat 3}-W
\eea
Note that $W$ is supercovariant (as $\psi_{a-}$ is supercovariant) under $\da^\rr(\ep_{+})$.
Note also that plugging $\zeta_{-}^\rr(\ep_{+})=-\half i\ga_5 \ep_{+} S_3$ into (\ref{la3ab}), we find
\bea
\la_3^{a b}=\xi^c\wh\om_c^{+}{}^{a b}+(\epbar_{2+}i\ga_5\ga^{a b}\ep_{1+})S_3
\eea

To establish the connection with the $d=3$ expressions as given in \cite{bvn1}, we now introduce a decomposition of 4-component spinors $\psi_{\pm}$ and $4\times 4$ gamma matrices $\ga^a,\ga^{\hat 3},\ga_5$ into 2-component spinors $\psi_{1,2}$ and $2\times 2$ gamma matrices $\wh\ga^a$:
\bea
\label{twocomp}
\psi=\bpm \psi_1 \\ \psi_2 \epm, \quad
\ga^{\hat 3}=\bpm 1 & 0 \\ 0 & -1 \epm \qrq
\psi_{+}=\bpm \psi_1 \\ 0 \epm, \quad
\psi_{-}=\bpm 0 \\ \psi_2 \epm
\nn\\
\ga^a=\bpm 0 & \wh\ga^a \\ \wh\ga^a & 0 \epm \qrq
\ga^{a b}=\bpm \wh\ga^{a b} & 0 \\ 0 & \wh\ga^{a b} \epm, \quad
i\ga_5=\bpm 0 & -1 \\ 1 & 0 \epm
\eea
where $\ga_5 \equiv \ga^{\hat 1}\ga^{\hat 2}\ga^{\hat 3} i\ga^{\hat 0}$ with $\ga_5^2=1$, and we require that $\wh\ga^{a b c}=+\ep^{a b c\hat 3}$. For Dirac conjugation we have
\bea
\psibar \equiv \psi^\dagger i\ga^{\hat 0}=(\psibar_2, \psibar_1), \quad
\psibar_{1,2} \equiv \psi_{1,2}^\dagger i\wh\ga^{\hat 0}
\eea
Using this decomposition with $\ep_{+}=\bpm \ep_1 \\ 0 \epm$, $\ep_{1+}=\bpm \ep_{11} \\ 0 \epm$ and $\ep_{2+}=\bpm \ep_{21} \\ 0 \epm$ we obtain
\bea
\da^\rr(\ep_1)e_m{}^a &=& \epbar_1\wh\ga^a\psi_{m1} \nn\\[2pt]
\da^\rr(\ep_{1})\psi_{m1} &=& 2D^\rr(\wh\om^{1})_m\ep_{1}+\half \wh\ga_m\ep_{1} S_3 \nn\\[2pt]
\la_3^{a b} &=& \xi^c\wh\om_c^{1}{}^{a b}+(\epbar_{21}\wh\ga^{a b}\ep_{11})S_3
\eea
where $\wh\om_{m a b}^{1}$ depends only on $e_m{}^a$ and $\psi_{m1}$. We observe that this is exactly the structure of $d=3$ expressions. Therefore, 
\bea
(e_m{}^a, \;\; \psi_{m1}, \;\; S_3)
\eea
is indeed the correct $d=3$ $N=1$ (Poincar\'e) sugra multiplet, and $\da^\rr(\ep_1)$ indeed closes into the standard $d=3$ $N=1$ (Poincar\'e) susy algebra,
\bea
[\da^\rr(\ep_{11}),\da^\rr(\ep_{21})]=\da_{g.c.}(\xi^a e_a{}^m)
+\da^\rr\Big( -\half\xi^a\psi_{a1} \Big)
+\da_L\Big( \xi^c\wh\om_c^{1}{}^{a b}+(\epbar_{21}\wh\ga^{a b}\ep_{11})S_3 \Big)
\eea
where $\xi^a=2(\epbar_{21}\wh\ga^a\ep_{11})$. 

For a (composite) $d=3$ $N=1$ scalar multiplet $(\wt A,\wt\chi,\wt F)$ on the boundary, we can now write a (separately $\da^\rr(\ep_1)$ supersymmetric) $F$-density
\bea
\wt{\mc{L}}_F=e_3\Big( \wt F+\half\psibar_{a1}\wh\ga^a\wt\chi
+\qter\wt A\psibar_{a1}\wh\ga^{a b}\psi_{b1}+\wt A S_3 \Big)
\eea
This boundary $F$-density formula, in conjunction with the ``$F+A$'' formula (\ref{FplusA}), provides means to construct general $\da^\rr(\ep_{+})$ supersymmetric bulk-plus-boundary actions:
\bea
S=\int_{\mc{M}} d^4x \mc{L}_F -\int_{\p\mc{M}} d^3x e_3 A +\int_{\p\mc{M}} d^3x \wt{\mc{L}}_F
\eea

\section{Extrinsic curvature multiplet in OMA sugra} 

The $U(1)_A$ compensator $\phi$ of OMA sugra has not appeared explicitly in the discussion of the induced supergravity multiplet. However, it becomes an essential part of the extrinsic curvature multiplet of OMA sugra, as we now demonstrate.

Although $\phi$ is a susy singlet under $\da_Q^P(\ep)$ transformation of OMA sugra, see (\ref{singlet}), it is not a singlet under the (modified) induced susy transformation $\da^\rr(\ep_{+})$
\bea
\da^\rr(\ep_{+})\phi=\ta(\ep_{+})=-\frac{2}{3}\epbar_{+}i\ga^5\ga^a\psi_{a-} \qrq
\da^\rr(\ep_{+})a=\epbar_{+}i\ga^5\ga^{a}\psi_{a-}
\eea
where $a=-\frac{3}{2}\phi$. On the other hand, from the result for $\da^\rr(\ep_{+})\psi_{a-}$ in (\ref{dapsiam}) and the explicit form for $\zeta_{+}(\ep_{+})$ in OMA sugra in (\ref{zetaspm}), we have
\bea
\da^\rr(\ep_{+})\psi_{a-}=\ga^b\ep_{+}\wh K_{a b}
-\frac{3}{2}i\ga_5\ep_{+}\wh A_a
+\frac{1}{3}\ga_a(S+i\ga_5\ga^b B_b)\ep_{+}
\eea
Contracting with $\ga^a$ and using $\ga^a\ga_a=3$ and $\ga^a\ga^b\wh K_{a b}=\eta^{a b}\wh K_{a b}\equiv\wh K$, we find
\bea
\da^\rr(\ep_{+})(\ga^a\psi_{a-})=(\wh K+S)\ep_{+}+i\ga_5\ga^a\ep_{+}(\frac{3}{2}\wh A_a+B_a)
\eea
According to (\ref{zetas}) and (\ref{scKA}), $B_a=-\frac{3}{2}A_a-\p_a a$ and $\wh A_a=A_a+\frac{1}{3}\psibar_{a+}i\ga_5\ga^b\psi_{b-}$, which gives
\bea
\frac{3}{2}\wh A_a+B_a=-\p_a a+\half\psibar_{a+}i\ga_5\ga^b\psi_{b-} =-\wh D_a a
\eea
where $\wh D_a a$ is the $\da^\rr(\ep_{+})$ supercovariant derivative of $a$. Therefore,
\bea
\da^\rr(\ep_{+})(\ga^a\psi_{a-})=(\wh K+S)\ep_{+}-i\ga_5\ga^a\ep_{+} \wh D_a a
\eea
Converting to the 2-component notation of (\ref{twocomp}), we finally find that
\bea
\da^\rr(\ep_1)a=\epbar_1\wh\ga^a\psi_{a2}, \quad
\da^\rr(\ep_1)(\wh\ga^a\psi_{a2})=(\wh K+S)\ep_1+\wh\ga^a\ep_1\wh D_a a
\eea
where $\wh D_a a=\p_a a-\half\psibar_{a1}\wh\ga^b\psi_{b2}$. This shows that 
\bea
(a, \;\; \wh\ga^a\psi_{a2}, \;\; \wh K+S) 
\eea
is a standard $d=3$ $N=1$ (Poincar\'e) scalar multiplet. As it contains the trace $\wh K$ of the (supercovariant) extrinsic curvature tensor $\wh K_{a b}$, we call it the extrinsic curvature multiplet. Therefore, we found that the $U(1)_A$ compensator $a$ (or $\phi$) plays a geometrical role: it is the first component of the extrinsic curvature multiplet.

\section{Conclusions} 

To summarize, we have extended the program of ``susy without BC'' \cite{bvn1} to $d=4$ $N=1$ Poincar\'e sugra. The new ingredient of the $d=4$ $N=1$ (superconformal) algebra compared to the one in the $d=3$ $N=1$ case considered in \cite{bvn1} is the $U(1)_A$ $R$-symmetry. We found that this symmetry plays a crucial role for the bulk-plus-boundary supersymmetry. The bulk sugra must have the local $U(1)_A$ among its symmetries for the program of ``susy without BC'' to work. This was demonstrated explicitly on the example of old-minimal (OM) sugra, where the $U(1)_A$ has been gauge fixed and correspondingly the ``$B$ problem'' arose in the attempt to make the bulk $F$-density supersymmetric in the presence of boundary. 

To resolve this problem with only a minor deviation from the OM sugra set of auxiliary fields, we have introduced a $U(1)_A$ compensator $\phi$ (or $a=-\frac{3}{2}\phi$) while at the same time restoring the role of $A_M^\text{aux}$ (or rather $A_M=-\frac{2}{3}A_M^\text{aux}$) as the $U(1)_A$ gauge field; we call this new version of supergravity the OMA sugra. 

Having restored the local $U(1)_A$, we managed to complete the ``susy without BC'' program. The resulting bulk-plus-boundary action formula is formally the same ``$F+A$'' formula as for the $d=3$ $N=1$ case, but the (modified) induced susy transformation $\da^\rr(\ep_{+})$ contains, in addition to $\da_Q^P(\ep_{+})$ and the (compensating) Lorentz transformation $\da_L(\la_{a\hat 3}(\ep_{+})=-\epbar_{+}\psi_{a-})$ (both present in the $d=3$ case), also a particular $\ep_{+}$ dependent $U(1)_A$ transformation, $\da_A(\ta(\ep_{+})=-\frac{2}{3}\epbar_{+}i\ga_5\ga^a\psi_{a-})$. The key check that $\da^\rr(\ep_{+})$ has been correctly identified came from showing that the commutator of two such transformations closes into the standard $d=3$ $N=1$ (Poincar\'e) susy algebra. This also allowed us to identify a subset of fields of bulk $d=4$ sugra as the fields of the standard $d=3$ sugra multiplet. This multiplet could then be straightforwardly used to construct separately susy boundary actions using the standard $d=3$ $F$-density formula. 

In addition to the induced sugra multiplet, we have identified the complementary extrinsic curvature tensor multiplet and discovered that the compensator $\phi$ (or rather $a$) is the first component in this multiplet. This is an example of a general phenomenon: certain pure gauge bulk degrees of freedom may turn into physically (or even geometrically) relevant fields in the presence of a boundary. (We observed another such example in the (rigidly susy) Chern-Simons theory in $d=3$ \cite{bvn2}.)

The analysis of new-minimal (NM) sugra, another Poincar\'e sugra with the local $U(1)_A$ preserved, was performed in parallel with that for OMA sugra. We intend to present several applications of our formalism in both OMA and NM sugra in a later publication.

Finally, in Appendix \ref{appA} we discussed how the same analysis can be performed without ever imposing the Lorentz $e_m{}^{\hat 3}=e_a{}^3=0$ gauge but using projection operators. These projection operators resemble the operators used in the derivation of the Gauss-Codazzi equations for induced curvatures, but we needed to extend this formulation to the case of vielbeins instead of metrics. We found that working with those projection operators gives results which are isomorphic to working in the gauge, and since the latter procedure is much simpler, we decided to use in the main text only the gauge-fixed approach.

\vspace{10pt}
\noindent
{\bf Acknowledgments.} This research was supported by the National Science Foundation (NSF) grant PHY-0653342. The research of D.B. was also supported in part by the German Science Foundation (DFG).

\appendix

\section{Projective formulation} 
\label{appA}

In this appendix we demonstrate how the modification of $\ep_{+}$ susy by an off-diagonal Lorentz transformation, see (\ref{modsusy}), arises in a geometrical approach (that we call ``projective formulation'') where, instead of imposing the gauge (\ref{egauge}), we work with projected indices and projected transformations. There is a simple correspondence between objects (such as the induced vielbein, etc.) in the ``gauge-fixed'' and ``projective'' formulations that will become clear as we proceed. Once this correspondence is established, the results in the two formulations become isomorphic.

The projective formulation is the covariant formulation where geometrical objects related to boundaries (or hypersurfaces) such as the induced metric (the first fundamental form), the extrinsic curvature (the second fundamental form), the induced covariant derivative, etc., are defined \cite{wald} using the projector $P_M{}^N=\da_M{}^N-n_M n^N$ where $n_M$ is the (unit, outward pointing) normal to the boundary. In applications to General Relativity it is sufficient to work with tensors having world indices ($M,N$); however, when fermions are present, one must also introduce tangent vector indices ($A,B$) and spinor indices ($\al,\beta$). The corresponding projective formulation has been developed and applied before (see e.g. \cite{LMoss}), but to the best of our knowledge the extension to vielbeins and projected susy transformations (see below) have not been discussed in the literature.

First of all, we note that using a projective formulation \emph{for world indices} is not needed for our purposes. We can freely choose our coordinates $x^M$ in such a way that the boundary $\p\mc{M}$ is at $x^3=0$ and the space $\mc{M}$ is ``to the right'' of $\p\mc{M}$ (i.e. $x^3>0$ for points in $\mc{M}$). This choice in no way restricts the local parameter $\xi^M(x)$ of general coordinate transformations. Making this choice, our normal $n_M$ and its tangent space analog $n_A=e_A{}^M n_M$ are given by
\bea
n_M=(\vec 0,-\frac{1}{\sqrt{g^{33}}}) \qrq
n_A=-\frac{e_A{}^3}{\sqrt{g^{33}}}
\eea
where the normalization follows from $g^{M N}n_M n_N=\eta^{A B}n_A n_B=1$ and the minus sign ensures that the normal is outward pointing. We define the following projectors for tangent vectors and spinors\footnote{
In local superspace there is a distinction between world space spinor indices and tangent space spinor indices. However, for spinors defined in $x$-space, one identifies the two concepts (by taking the ``spinor vielbein'' to be unity) and speaks simply of ``spinor indices.''}
\bea
P_A{}^B=\da_A{}^B-n_A n^B, \quad N_A{}^B=n_A n^B, \quad
P_{\pm}=\half(1\pm n_A\ga^A)
\eea
where the spinor indices of $P_{\pm}$ have been suppressed. The projectors satisfy the standard properties ($P_A{}^B+N_A{}^B=\da_A{}^B$, $P_{+}+P_{-}=1$, $P_A{}^B P_B{}^C=P_A{}^C$, $P_A{}^B N_B{}^A=0$, $P_{+}P_{-}=0$, etc.) and allow a decomposition of Lorentz vectors $V_A$ and spinors $\psi$ into ``parallel'' ($V_{A^\rr},\psi_{+}$) and ``normal'' ($V_{\dot A},\psi_{-}$) components
\bea
V_A=V_{A^\rr}+V_{\dot A}, \quad 
V_{A^\rr} \equiv P_A{}^B V_B, \quad
V_{\dot A} \equiv N_A{}^B V_B; \quad
\psi=\psi_{+}+\psi_{-}, \quad 
\psi_{\pm} \equiv P_{\pm}\psi
\eea
We also define $V_{\dot 3} \equiv n^A V_A$ so that $V_{\dot A}=n_A V_{\dot 3}$.

Applying this formalism to the Lorentz index of the vielbein $e_M{}^A=(e_m{}^A,e_3{}^A)$ and its inverse $e_A{}^M=(e_A{}^m,e_A{}^3)$, we immediately find two \emph{identities}
\bea
\label{ident}
e_m{}^{\dot 3} &\equiv& e_m{}^A n_A =-\frac{e_m{}^A e_A{}^3}{\sqrt{g^{33}}}=0 \nn\\
e_{A^\rr}{}^3 &\equiv& P_A{}^B e_B{}^3=-\sqrt{g^{33}} P_A{}^B n_B=0
\eea
Defining $\wh g_{m n} \equiv e_m{}^{A^\rr} e_n{}^{B^\rr}\eta_{A B}$ and $\wh g^{m n} \equiv \eta^{A B}e_{A^\rr}{}^m e_{B^\rr}{}^n$ where
\bea
e_m{}^{A^\rr} \equiv e_m{}^B P_B{}^A, \quad
e_{A^\rr}{}^m \equiv P_A{}^B e_B{}^m
\eea
a short calculation shows that $\wh g_{m n}=e_m{}^A e_n{}^B\eta_{A B}=g_{m n}$ which is the induced metric at a hypersurface with constant $x^3$, whereas $\wh g^{m n}=P^{A B}e_A{}^m e_B{}^n=g^{m n}-(g^{m3}g^{n3}/g^{33})$ satisfies $\wh g_{m k} \wh g^{k n}=\da_m{}^n$ and is thus the inverse of the induced metric. In this sense $e_m{}^{A^\rr}$ can be called the induced vielbein and $e_{A^\rr}{}^m$ its inverse (although both are not even square matrices).\footnote{
If one does not make the $M=(m,3)$ decomposition, but uses both world space and tangent space projectors, the induced vielbein is $P_M{}^N e_N{}^B P_B{}^A$. This is a square matrix but with vanishing determinant.}

In General Relativity, given a bulk (world space) tensor $T$ and the bulk covariant derivative $\nabla_M$, we have \emph{two} candidates for the induced (``hypersurface compatible'') covariant derivative of the corresponding projected tensor $T^\rr \equiv P T$,
\bea
\nabla^\rr T^\rr \equiv P \nabla T, \quad
\nabla^{\rr\rr} T^\rr \equiv P \nabla T^\rr
\eea
where ``$P$'' is a symbolic projector whose precise form depends on the index structure of the tensor to which it is applied. For $T^\rr_M \equiv T_{M^\rr}\equiv P_M{}^N T_N$, we have
\bea
\nabla_M^\rr T_{N^\rr} &\equiv& P_N{}^{N_1} P_M{}^{M_1} \nabla_{M_1} T_{N_1} \nn\\
\nabla_M^{\rr\rr} T_{N^\rr} &\equiv& P_N{}^{N_1} P_M{}^{M_1} \nabla_{M_1} (P_{N_1}{}^{N_2} T_{N_2})
\eea
Defining the extrinsic curvature tensor as $K_{M N}=-P_M{}^{M_1} P_N{}^{N_1} \nabla_{N_1} n_{M_1}=K_{N M}$ (see \cite{wald,bvn1}), we find a relation between both derivatives
\bea
\nabla_M^{\rr\rr} T_{N^\rr}=\nabla_M^\rr T_{N^\rr}+K_{M N} (n^K T_K)
\eea
We used $P_M{}^{M_1}P_N{}^{N_1} \nabla_{N_1} P_{M_1}{}^K=K_{M N} n^K$, which in turn implies
\bea
\nabla_M^\rr P_N{}^K \equiv P_N{}^{N_1} P_M{}^{M_1} (\nabla_{M_1} P_{N_1}{}^{K_1})P_{K_1}{}^K=0
\eea
Thus the projector $P_M{}^N$ commutes with the projected derivative $\nabla^\rr$. Since $\nabla^\rr=\nabla^{\rr\rr}$ on the projector, it commutes with both $\nabla^\rr$ and $\nabla^{\rr\rr}$: $\nabla^\rr P=\nabla^{\rr\rr}P=0$. Similarly, if the original tensor $T$ is already projected, $T=P T$, then $\nabla^\rr T=\nabla^{\rr\rr}T$ (which may be the reason why $\nabla^\rr$ and $\nabla^{\rr\rr}$ are usually not distinguished \cite{wald}).

We now use this approach to define projected transformations. (Recall that we use only projectors for tangent space vector and spinor indices; world space vector indices are simply decomposed as $M=(m,3)$.) Given a transformation $\da$, defined for a bulk (Lorentz) tensor $T$, we define, for a projected tensor $T^\rr \equiv P T$, the corresponding projected transformation $\da^\rr$ as follows
\bea
\label{projtr}
\da^\rr T^\rr \equiv P\da T 
\eea
where $P$ is the identity for scalars, or any of $P_A{}^B, N_A{}^B, P_{\pm}$ or their tensor products for tensors and spinors. For example, $\da^\rr T_{A^\rr \dot B}=P_A{}^{A_1} N_B{}^{B_1} \da T_{A_1 B_1}$. It immediately follows that for \emph{any} transformation $\da$, the corresponding projected transformation $\da^\rr$ of the normal vector $n_A=N_A{}^B n_B$ vanishes
\bea
\da^\rr n_A \equiv N_A{}^B\da n_B=0
\eea
where we used $n^A\da n_A=0$ which follows from the normalization condition $n^A n_A=1$. One can similarly prove that all projectors are invariant under (i.e. commute with) the projected transformations
\bea
\label{invproj}
\da^\rr (P_A{}^B, N_A{}^B, P_{\pm})=0
\eea
For example, $\da^\rr P_{+}=P_{+}(\da P_{+})P_{+}=\half(\da n_A)(P_{+}\ga^A P_{+})=\half (n^A\da n_A)P_{+}=0$ where we used $n^A\da n_A=0$ and the identity $P_{\pm}\ga^A P_{\pm}=\pm n^A P_{\pm}$ which follows from $P_{\pm}\ga^A=\ga^A P_{\mp}\pm n^A$.

For projected Lorentz transformations we have, according to our definition in (\ref{projtr}), $\da^\rr(\la)V_{A^\rr}=P_A{}^B\da(\la)V_B=P_A{}^B\la_B{}^C V_C$, etc. It is straightforward to show that
\bea
\label{projLorentz}
&& \da^\rr(\la)V_{A^\rr}=\la_{A^\rr}{}^{B^\rr} V_{B^\rr}+\la_{A^\rr}{}^{\dot 3} V_{\dot 3}, \quad
\da^\rr(\la)V_{\dot 3}=\la_{\dot 3}{}^{B^\rr} V_{B^\rr} \nn\\
&& \da^\rr(\la)\psi_{\pm}=\qter\la^{A^\rr B^\rr}\ga_{A^\rr B^\rr}\psi_{\pm}
+\half\la^{A^\rr\dot 3}\ga_{A^\rr\dot 3}\psi_{\mp}
\eea
where $\ga_{A^\rr}=P_A{}^B\ga_B$ and $\ga_{\dot 3}=n^A\ga_A$. As a consequence of $\{ \ga_A,\ga_B \}=2\eta_{A B}$, we find that
\bea
\{ \ga^{A^\rr},\ga^{B^\rr} \}=P^{A B}, \quad
\{ \ga^{A^\rr},\ga^{\dot 3} \}=0, \quad
\{ \ga^{\dot 3},\ga^{\dot 3} \}=2
\eea
(Note that $V^{A^\rr}\equiv V^B P_B{}^A=P^{A B}V_B=\eta^{A B}V_{B^\rr}$ while $V^{\dot 3} \equiv n_A V^A=n^A V_A \equiv V_{\dot 3}$.)

Let us now turn to the projected supersymmetry transformations. We define parameters $\ep_{\pm}$ by $\ep_{\pm}=P_{\pm}\ep$ which yields $\ga^{\dot 3}\ep_{\pm}=\pm\ep_{\pm}$. Note that $\ep_{\pm}$ are field-dependent.
Starting from $\da(\ep)e_M{}^A=\epbar\ga^A\psi_M$, we find for $\da^\rr(\ep_{+})\equiv P\da(\ep_{+})$ acting on the projected parts of the vielbein $e_M{}^A$ the following results
\bea
\ba[b]{rclcrcl}
\da^\rr(\ep_{+})e_m{}^{A^\rr} &=& \epbar_{+}\ga^{A^\rr}\psi_{m+}, &\quad&
\da^\rr(\ep_{+})e_m{}^{\dot 3} &=& -\epbar_{+}\psi_{m-} \\[3pt]
\da^\rr(\ep_{+})e_3{}^{A^\rr} &=& \epbar_{+}\ga^{A^\rr}\psi_{3+}, &\quad&
\da^\rr(\ep_{+})e_3{}^{\dot 3} &=& -\epbar_{+}\psi_{3-}
\ea
\eea
At this point we note that we have run into a problem: our projected susy transformation does not preserve the identity $e_m{}^{\dot 3} \equiv e_m{}^A n_A \equiv 0$ of (\ref{ident})! On the other hand, it is still true that $\da(\ep_{+})e_m{}^{\dot 3}=0$. It is easy to understand what is going on from the following identity:
\bea
\da^\rr e_m{}^{\dot 3}=n_A\da e_m{}^A=\da e_m{}^{\dot 3}-e_m{}^{A}\da n_A
=\da e_m{}^{\dot 3}-e_m{}^{A^\rr}\da n_A
\eea
For a general variation $\da$ of $n_A=-e_A{}^3/\sqrt{g^{33}}$ one finds
\bea
\da n_A=-\frac{P_A{}^B\da e_B{}^3}{\sqrt{g^{33}}}
\eea
which gives $\da(\ep)n_A=-\epbar\ga^{\dot 3}\psi_{A^\rr}$ and therefore $\da(\ep_{+})n_A=\epbar_{+}\psi_{A^\rr-}$. (Note that this is consistent with $\da^\rr(\ep_{+})n_A=N_A{}^B\da(\ep_{+})n_B=0$.) Now it is clear that $\da^\rr(\ep_{+})e_m{}^{\dot 3}\neq 0$ is due to $\da(\ep_{+})n_A\neq 0$ even though $\da(\ep_{+})e_m{}^{\dot 3}=0$.

The identity $e_m{}^{\dot 3} \equiv 0$ is also not preserved by the projected Lorentz transformations as $\da^\rr(\la)e_m{}^{\dot 3}=-e_m{}^{A^\rr}\la_{A^\rr}{}^{\dot 3}$. For a combined transformation $\da^\rr(\ep,\la)=\da^\rr(\ep)+\da^\rr(\la)$ we find
\bea
\da^\rr(\ep,\la)e_m{}^{\dot 3}=-e_m{}^{A^\rr} \Big( \la_{A^\rr \dot 3}+\da(\ep)n_A \Big)
\eea
Therefore, preservation of $e_m{}^{\dot 3}\equiv 0$ forces us to modify the projected transformations by adding a compensating Lorentz transformation with parameter $\la_{A^\rr\dot 3}=-\da n_A$. For $\ep_{+}$ susy this leads to the following modified projected transformation
\bea
\da^{\rr\rr}(\ep_{+})=\da^\rr(\ep_{+})
+\da_L^\rr\Big( \la_{A^\rr 3}(\ep_{+})=-\epbar_{+}\psi_{A^\rr-} \Big)
\eea
Similar modifications are required for all other projected transformations. Since under Lorentz transformations $\da(\la)n_A=\la_A{}^B n_B=\la_{A^\rr}{}^{\dot 3}$, we find that
\bea
\da^{\rr\rr}(\la_{A^\rr B^\rr})=\da^\rr(\la_{A^\rr B^\rr}), \quad
\da^{\rr\rr}(\la_{A^\rr \dot 3})=0
\eea
whereas for general coordinate transformations it follows from $\da(\xi)n_A=\xi^M\p_M n_A$ that\footnote{
When one imposes the gauge $e_m{}^{\hat 3}=e_a{}^3=0$, one finds $n_A=(0,0,0,-1)$. Then $\da(\ep_{+})n_A=0$, but one needs a compensating Lorentz transformation to stay in the gauge, and the final result for the modified $\ep_{+}$ transformation has the same (or, rather, isomorphic) form in both approaches. On the other hand, we needed no modification for the $\xi^m$ part of the general coordinate transformation in the gauge-fixed case \cite{bvn1} which is also in accord with (\ref{modgc}) since $\xi^m\p_m n_A=0$ in the gauge-fixed case.}
\bea
\label{modgc}
\da^{\rr\rr}(\xi)=\da^\rr(\xi)+\da_L^\rr\Big( \la_{A^\rr \dot 3}=-\xi^M\p_M n_A \Big)
\eea

Having come so far, let us ask what happens if, instead of $\da^\rr T^\rr \equiv P\da T$, as in (\ref{projtr}), we define an (alternative) projected transformation by
\bea
\label{modprojtr}
\da^{\rr\rr} T^\rr \equiv P\da T^\rr 
\eea
where $T^\rr \equiv P T$. Obviously, we have $\da^{\rr\rr}T^\rr=\da^\rr T^\rr+(P\da P)T$. (Note that $\da^\rr P=P(\da P)P$ in (\ref{invproj}) vanishes, but, as we shall show, $P\da P$ is nonvanishing.) Writing this out more explicitly for the basic projected tensors and spinors $T^\rr=(V_{A^\rr},V_{\dot A},\psi_{\pm})$, we find
\bea
\da^{\rr\rr}V_{A^\rr} &=& \da^\rr V_{A^\rr}+(P_A{}^B\da P_B{}^C) V_C, \quad
P_A{}^B\da P_B{}^C=-(P_A{}^B\da n_B) n^C=-(\da n_A)n^C \nn\\[5pt]
\da^{\rr\rr}V_{\dot A} &=& \da^\rr V_{\dot A}+(N_A{}^B\da N_B{}^C)V_C, \quad
N_A{}^B\da N_B{}^C=n_A n^B n_B\da n^C =n_A\da n^C \nn\\
\da^{\rr\rr}\psi_{\pm} &=& \da^\rr\psi_{\pm}+(P_{\pm}\da P_{\pm})\psi, \quad
P_{\pm}\da P_{\pm}=P_{\pm}(\pm\half\da n_A \ga^A)
=-\half\da n_A\ga^A\ga^{\hat 3}P_{\mp}
\eea
where we used $P_A{}^B n_B=0$, $n^A\da n_A=0$, $n^B n_B=1$, $P_{\pm}\ga^A=\ga^A P_{\mp}\pm n^A$ and $\ga^{\dot 3}P_{\mp}=\mp P_{\mp}$. Comparing these results with the $\la_{A^\rr\dot 3}$ parts of projected Lorentz transformations in (\ref{projLorentz}), we find that, in all cases,
\bea
\da^{\rr\rr} T^\rr=\da^\rr T^\rr+\da_L^\rr(\la_{A^\rr\dot 3}=-\da n_A)
\eea
This shows that the modified projected transformations are precisely the projected transformations defined by (\ref{modprojtr}). (Note also that $\da^{\rr\rr} T^\rr=P\da T^\rr$ is in line with the definition of the induced covariant derivative: $\nabla^{\rr\rr} T^\rr=P\nabla T^\rr$.) 

Calculating variations of the bulk fields $e_M{}^A$, $\psi_M$, etc. under the modified projected susy transformation $\da^{\rr\rr}(\ep_{+})$, we observe that they have the same form as that found in the gauge $e_m{}^{\hat 3}=e_a{}^3=0$ provided we make the identification $(A^\rr,\dot 3)\lra (a,\hat 3)$. 
(Actually, $V^{\dot 3}\equiv n_A V^A$ becomes $-V^{\hat 3}$ in the gauge, but this minus sign can be removed by redefining $V^{\dot 3}$.)
It is then (almost) obvious that for the commutator of two modified projected susy transformations we find a result isomorphic to the result in the gauge, with all the transformations on the right hand side being again the modified projected transformations $\da^{\rr\rr}$. For example, in the $d=3$ $N=1$ case of \cite{bvn1}, where only the Lorentz modification was required for induced $\ep_{+}$ susy, the susy algebra in the projective formulation has the following form,
\bea
[\da^{\rr\rr}(\ep_{1+}),\da^{\rr\rr}(\ep_{2+})]=\da^{\rr\rr}_{g.c.}(\xi^m)
+\da^{\rr\rr}(\ep_{+})
+\da_L^{\rr\rr}(\la^{A^\rr B^\rr})
\eea
with $\xi^m=2(\epbar_{2+}\ga^{A^\rr}\ep_{1+})e_{A^\rr}{}^m$, $\ep_{+}=-\half\xi^m\psi_{m+}$ and $\la_{A^\rr B^\rr}=\xi^m\wh\om_{m A^\rr B^\rr}^{+}$. 
This is the same form as obtained in \cite{bvn1} in the Lorentz gauge.
One subtlety to be clarified is whether the field dependence of susy parameters plays any role in obtaining this result. A priori one could expect contributions to the composite parameter of the modified susy transformation which stem from the field dependence of $\ep_{+}\equiv\half(1+n_A\ga^A)\ep$
\bea
\wt\ep=\da^{\rr\rr}(\ep_{1+})\ep_{2+}-(1\lra 2)
\eea
However, if $\ep$ itself is field independent, then all the field dependence in $\ep_{+}$ is due to the projector $P_{+}$. Since, as we showed, the projectors are invariant under arbitrary projected transformations, we find that $\wt\ep=0$. Note, however, that for this argument to be correct we should never require that $\ep_{-}\equiv\half(1-n_A\ga^A)\ep$ vanishes as this would violate the assumption that $\ep$ is field independent. We simply concentrate on susy transformations with $\ep_{+}$, leaving $\ep_{-}$ aside.

Another subtlety in lifting the results found in the gauge $e_m{}^{\hat 3}=e_a{}^3=0$ to the corresponding results in the projective formulation has to do with the determinant of the induced vielbein ($\wh e_2$ in the $d=3$ case), which we definitely cannot define as the determinant of $e_m{}^{A^\rr}$. Instead we define
\bea
\wh e_2=\sqrt{-\det \wh g_{m n}}, \quad
\wh g_{m n}=e_m{}^{A^\rr}e_n{}^{B^\rr}\eta_{A B}
\eea
This definition gives $\da\wh e_2=\half\wh e_2\wh g^{m n}\da\wh g_{m n}=\wh e_2 e_{A^\rr}{}^m\da e_m{}^{A^\rr}$ which coincides with the lifting of $\da e_2=e_2 e_a{}^m\da e_m{}^a$ in the gauge-fixed case (where $e_2=\det e_m{}^a$). Therefore, the $d=3$ ``$F+A$'' formula in the projective formulation is
\bea
S=\int_{\mc{M}} d^3x \mc{L}_F-\int_{\p\mc{M}} d^2x \wh e_2 A
\eea

We conclude that as far as Lorentz modification of induced transformation rules is concerned, all results in all dimensions in the gauge $e_m{}^{\hat 3}=e_a{}^3=0$ can be recast in the language of the projective formulation. In the case of local susy in $d=3$ there is only a Lorentz modification, but in the case of local susy in $d=4$ one needs also a $U(1)_A$ modification to obtain closure of the gauge algebra. This $U(1)_A$ modification is added after one has deduced the Lorentz modification as discussed in this appendix.

\section{Old-minimal sugra with a $\bf U(1)_A$ compensator} 
\label{appB}

The derivation of old-minimal $d=4$ $N=1$ Poincar\'e sugra from conformal sugra was performed in \cite{KT} and summarized in \cite{PVN,KU}. The gauge fields and symmetry parameters in the conventions of \cite{KU} are
\bea
h_\mu &=& e_\mu{}^m P_m+\half\om_\mu{}^{m n} M_{m n}
+\psibar_\mu Q+f_\mu{}^m K_m+b_\mu D+\phibar_\mu S+A_\mu A \nn\\
\eps &=& \xi^m P_m+\half\la^{m n}M_{m n}
+\epbar Q+\xi_K^m K_m+\la_D D+\zetabar S+\ta A
\eea
(We adhere to these conventions in this appendix; changing $(\mu,m)\rightarrow(M,A)$ and $\ep\rightarrow 2\ep$ brings us to the conventions used in the main text.) 

To derive old-minimal sugra with a $U(1)_A$ compensator (OMA sugra, for short) we will follow the standard derivation with one small (but essential) difference: we will not impose the $U(1)_A$ gauge condition. We start with the conformal sugra multiplet $(e_\mu{}^m,\psi_\mu,A_\mu,b_\mu)$ and a chiral multiplet\footnote{
Usually one calls the operator $\half(1+\ga_5)$ the projection operator onto left-handed fermions, but because in \cite{KU} it is denoted by $P_R$, we will also here denote it by $P_R$.}
\bea
(\mc{A}_0, \;\; P_R\chi_0, \;\; \mc{F}_0)=\Big(
\half(A_0+i B_0), \;\; \half(1+\ga_5)\chi_0, \;\; \half(F_0+i G_0) \Big)
\eea
of $U(1)_A$ weight $n=1$. Note that under $U(1)_A$ transformations with local parameter $\ta$ the supergravity fields and those of a chiral multiplet $(\mc{A},\chi_R \equiv P_R\chi,\mc{F})$ with $U(1)_A$ weight $n$ transform as follows
\bea
&& \da e_\mu{}^m=0, \quad
\da\psi_\mu=\frac{3}{4}i\ga_5\psi_\mu \ta, \quad
\da A_\mu=\p_\mu\ta, \quad
\da b_\mu=0 \nn\\[5pt]
&& \da\mc{A}=\frac{i}{2}n\ta\mc{A}, \quad
\da\chi_R=i(\frac{n}{2}-\frac{3}{4})\ta\chi_R, \quad
\da\mc{F}=\frac{i}{2}(n-3)\ta\mc{F} 
\eea
We consider $D$ (dilatations), $S$ (conformal supersymmetry) and $K_m$ (special conformal transformations) as extraneous symmetries and gauge fix them by setting
\bea
\label{constr}
2\mc{A}_0=e^{i\phi/2}, \quad \chi_0=0, \quad b_\mu=0
\eea
These constraints are invariant under general coordinate ($\da_{g.c.}(\xi^\mu)$) and local Lorentz ($\da_L(\la^{m n})$) transformations provided $\phi$ is a scalar.
The combined $Q,S,K_m,D$ and $A$ transformation of these constraints requires
\bea
\da\mc{A}_0 &\equiv&
\half\epbar\chi_{0 R}+\la_D \mc{A}_0+\frac{i}{2}\ta\mc{A}_0 
= \frac{i}{2}\mc{A}_0\da\phi \nn\\
\da\chi_{0 R} &\equiv&
P_R(\slD^c\mc{A}_0+\mc{F}_0)\ep+2\mc{A}_0\zeta_R+\frac{3}{2}\la_D\chi_{0 R}-\frac{i}{4}\ta\chi_{0 R}
=0 \nn\\
\da b_\mu &\equiv&
\half\epbar\phi_\mu-\half\zetabar\psi_\mu-2\xi_K^m e_{\mu m}+\p_\mu\la_D
=0
\eea
where $D_m^c \mc{A}_0=\p_m\mc{A}_0-\half\psibar_m\chi_{0 R}-b_m\mc{A}_0-\frac{i}{2}A_m\mc{A}_0$. This is solved by requiring
\bea
\label{secconstr}
\da\phi=\ta, \quad
\la_D=0, \quad
\xi_m^K=\qter(\epbar\phi_m-\zetabar\psi_m), \quad
2\zeta_R=-\mc{A}_0^{-1} P_R (\slD^c\mc{A}_0+\mc{F}_0)\ep
\eea
This tells us that the $U(1)_A$ symmetry is preserved provided it acts on $\phi$ with a shift, $\da_A(\ta)\phi=\ta$; the $D$ symmetry is broken and can be simply dropped; $S$ and $K_m$ symmetries are broken but play a role in restoring the $Q$ symmetry. The Poincar\'e susy is given by the ``$Q+S+K$'' formula \cite{PVN,KU}
\bea
\label{QSK1}
\da_Q^P(\ep) \equiv \da_Q(\ep)+\da_S(\zeta(\ep))+\da_K(\xi_K^m(\ep))
\eea
where $\zeta(\ep)$ and $\xi_K^m(\ep)$ are given in (\ref{secconstr}). For $\zeta(\ep)$ we find
\bea
P_R\zeta=-\half P_R(\wt{\mc{F}}_0-\frac{i}{2}\ga^m\wt A_m)\ep, \quad
\wt{\mc{F}}_0 \equiv \mc{A}_0^{-1}\mc{F}_0, \quad
\wt A_\mu \equiv A_\mu-\p_\mu\phi
\eea
where we note that $\da_A(\ta)\wt{\mc{F}}_0=-\frac{3}{2}i\ta\wt{\mc{F}}_0$ and $\da_A(\ta)\wt A_m=0$. Defining $\wt{\mc{F}}_0=\half(\wt F_0+i\wt G_0)$ and extracting the projector $P_R=\half(1+\ga_5)$, we obtain
\bea
\zeta=-\qter(\wt F_0+i\ga_5\wt G_0-i\ga_5\ga^m\wt A_m)\ep
\eea
Finally, for comparison with the conventional old-minimal (OM) formulation, we define\footnote{
With $2\mc{A}_0=e^{i\phi}$, we have $\wt{\mc{F}}_0=2 e^{-i\phi}\mc{F}_0$ or $\wt F_0+i\wt G_0=2 e^{-i\phi}(F_0+i G_0)$. For OM sugra with $\phi=0$, this gives $S=3 F_0$ and $P=-3 G_0$ \cite{KT,PVN,KU}.}
\bea
S=\frac{3}{2}\wt F_0, \quad
P=-\frac{3}{2}\wt G_0, \quad
A_\mu^\text{aux}=-\frac{3}{2}A_\mu, \quad
a=-\frac{3}{2}\phi
\eea
which gives for our OMA sugra
\bea
\label{OMAzeta1}
\zeta(\ep)=\half H\ep, \quad
H \equiv -\frac{1}{3}(S-i\ga_5 P+i\ga_5\ga^m B_m), \quad
B_\mu \equiv A_\mu^\text{aux}-\p_\mu a
\eea
This is the key formula that we need. Using the ``$Q+S+K$'' rule (taking into account that all independent fields are inert under $K$), it is straightforward to write explicitly Poincar\'e susy transformations of fields in the OMA sugra multiplet $(e_\mu{}^m,\psi_\mu,A_\mu^\text{aux},S,P,a)$, and fields in other multiplets (chiral, linear, vector, etc.).

From (\ref{secconstr}) we observe that $\phi$ (or $a$) shifts under $U(1)_A$, but is inert under $Q,S,K$. Therefore, it is inert under Poincar\'e susy. To understand how this can be consistent with the usual statement that ``two susy transformations yield a translation,'' we need to find the susy algebra for OMA sugra.

The commutator of two Poincar\'e susy transformations follows from the superconformal algebra and the ``$Q+S+K$'' rule \cite{PVN,KU} and we find,\footnote{
The $K$ transformation with parameter $\xi_m^K(\ep)=\qter(\epbar\phi_m-\zetabar\psi_m)$, as well as the field dependence of both $\zeta(\ep)$ and $\xi_m^K(\ep)$, are crucial for the recombination of composite $Q,S,K$ transformations on the right hand side of the commutator into the composite Poincar\'e susy transformation $\da_Q^P(-\xi^\mu\psi_\mu)$.}
for OM, OMA and NM sugra,
\bea
\label{QPQP1}
[\da_Q^P(\ep_1),\da_Q^P(\ep_2)] &=& \da_{g.c.}(\xi^\mu)+\da_Q^P(-\xi^\mu\psi_\mu)
+\da_A(-\xi^\mu A_\mu+\epbar_{[1}i\ga_5\zeta_{2]}) \nn\\
&& \hspace{115pt}
+\da_L(\xi^\mu\wh\om_\mu{}^{m n}+\half\epbar_{[1}\ga^{m n}\zeta_{2]})
\eea
where $\xi^\mu=\half\epbar_2\ga^\mu\ep_1$, $\zeta_{1,2}=\zeta(\ep_{1,2})$ and $\wh\om_\mu{}^{m n}$ is the usual supercovariant spin connection; we also introduced the notation $[1 2]=1 2 - 2 1$. Substituting $\zeta(\ep)$ of OMA sugra, see (\ref{OMAzeta1}), we find
\bea
[\da_Q^P(\ep_1),\da_Q^P(\ep_2)] &\overset{\text{OMA}}{=}& 
\da_{g.c.}(\xi^\mu)+\da_Q^P(-\xi^\mu\psi_\mu)
+\da_A(-\xi^\mu \p_\mu\phi) \nn\\
&& 
+\da_L(\xi^\mu\wh\om_\mu{}^{m n}
+\frac{1}{6}\epbar_2\ga^{m n}(S-i\ga_5 P)\ep_1
+\frac{1}{3}\ep^{m n \mu \nu}\xi_\mu B_\nu)
\eea
The form of the composite $U(1)_A$ transformation explains why $\phi$ can be inert under susy: the commutator of two Poincar\'e susy transformations on $\phi$ vanishes because the sum of the composite $\da_{g.c.}$ and $\da_A$ transformations of $\phi$ vanishes. Setting $\phi=0$ gives the algebra for OM sugra.

\section{Conventions and technical details} 
\label{appC}

Our conventions are the same as in \cite{bvn1} with the obvious extension from $d=3$ to $d=4$. $(M,N,K)$ are $d=4$ world (curved) indices; $(A,B,C)$ are $d=4$ tangent (flat) indices; spinor indices are always implicit. We use the decomposition $M=(m,3)$ and $A=(a,\hat 3)$ with $m=0,1,2$ and $a=\hat 0,\hat 1,\hat 2$. The space $\mc{M}$ has boundary $\p\mc{M}$ at $x^3=0$ with coordinates $x^m$; points in $\mc{M}$ have $x^3>0$. The gamma matrices $\ga^A, \ga_5$ satisfy
\bea
\ga^A\ga^B=\eta^{A B}+\ga^{A B}, \quad 
\ga_5=\ga^{\hat 1}\ga^{\hat 2}\ga^{\hat 3} i\ga^{\hat 0}, \quad
\ga^{A B C D}=i\ga_5\ep^{A B C D}
\eea
where $\eta^{A B}=(- + + +)$ and $\ep^{\hat 0\hat 1\hat 2\hat 3}=+1$. Our spinors are Majorana, $\psibar \equiv \psi^\dagger i\ga^{\hat 0}=\psi^T C$, where $(\ga^{\hat 0})^\dagger=-\ga^{\hat 0}$, $(\ga^{\hat 1},\ga^{\hat 2},\ga^{\hat 3})^\dagger=(\ga^{\hat 1},\ga^{\hat 2},\ga^{\hat 3})$ and $C^T=-C$, $C\ga^A C^{-1}=-(\ga^A)^T$. The spinorial projectors $P_{\pm}$ satisfy
\bea
P_{\pm} \equiv \half(1\pm\ga^{\hat 3}), \quad
P_{+}+P_{-}=1, \quad
P_{\pm}P_{\pm}=P_{\pm}, \quad
P_{+}P_{-}=0
\eea
where we stress that $\ga^{\hat 3}$ is constant. We decompose spinors as $\psi=\psi_{+}+\psi_{-}$ where $\psi_{\pm}=P_{\pm}\psi$. It follows that $\psibar_{\pm}=\psibar P_{\mp}$. Therefore, for example, $\phibar\psi=\phibar_{+}\psi_{-}+\phibar_{-}\psi_{+}$.

General coordinate $\da_{g.c.}(\xi)$ and local Lorentz transformations $\da_L(\la)$ of the vielbein $e_M{}^A$ and gravitino $\psi_M$ are given by
\bea
&& \da_{g.c.}(\xi)e_M{}^A=\xi^N\p_N e_M{}^A+(\p_M\xi^N)e_N{}^A, \quad
\da_L(\la)e_M{}^A=\la^{A B} e_{M B} \nn\\
&& \da_{g.c.}(\xi)\psi_M=\xi^N\p_N \psi_M+(\p_M\xi^N)\psi_N, \quad
\da_L(\la)\psi_M=\qter\la^{A B}\ga_{A B}\psi_M
\eea
with $\la^{A B}=-\la^{B A}$. We use $D(\om)_M$ to denote a Lorentz covariant derivative constructed with connection $\om_{M A B}$; for example, $D(\wh\om)_M\psi_N=\p_M\psi_N+\qter\wh\om_{M A B}\ga^{A B}\psi_N$. The supercovariant spin connection $\wh\om_{M A B}$ is given by
\bea
\label{CMNA}
&& \wh\om_{M A B}=\om(e)_{M A B}+\kappa_{M A B}, \quad
\kappa_{M A B}=\qter(\psibar_M\ga_A\psi_B-\psibar_M\ga_B\psi_A+\psibar_A\ga_M\psi_B) \nn\\
&& \om(e)_{M A B}=\half(C_{M A B}-C_{M B A}-C_{A B M}), \quad
C_{M N}{}^A=\p_M e_N{}^A-\p_N e_M{}^A
\eea
where $\psi_A=e_A{}^M\psi_M$, etc.; $\om(e)_{M A B}$ is the torsion-free connection and $\kappa_{M A B}$ is the contorsion tensor. Under local Lorentz transformations $\da(\la)\wh\om_{M A B}=-D(\wh\om)_M\la_{A B}$. 

The induced metric on a hypersurface with constant $x^3$ is $g_{m n}=e_m{}^a e_{n a}+e_m{}^{\hat 3} e_{n 3}$. In general, therefore, $e_m{}^a$ is not the induced vielbein. In the gauge $e_m{}^{\hat 3}=0$, however, $g_{m n}=e_m{}^a e_{n a}$ and $e_m{}^a$ is the induced vielbein. Imposing $e_m{}^{\hat 3}=0$ implies $e_a{}^3=0$, and vice versa. In the gauge $e_m{}^{\hat 3}=e_a{}^3=0$, we have $e_m{}^a e_a{}^m=\da_m{}^n$, $e_a{}^m e_m{}^b=\da_a{}^b$, $e_3{}^{\hat 3} e_{\hat 3}{}^3=1$ as well as
\bea
\ga_m=e_m{}^a\ga_a, \quad
\ga_3=e_3{}^a\ga_a+e_3{}^{\hat 3}\ga_{\hat 3}, \quad
\ga^m=\ga^a e_a{}^m+\ga^{\hat 3} e_{\hat 3}{}^m, \quad
\ga^3=\ga^{\hat 3} e_{\hat 3}{}^3
\eea
In addition, $\om(e)_{m a b}$ coincides with the torsion-free connection constructed out of $e_m{}^a$ whereas $K_{m n}=\om(e)_{m a\hat 3} e_n{}^a$ is the extrinsic curvature tensor \cite{bvn1}. Note that under local Lorentz transformations $\da(\la)\wh\om_{m a\hat 3}=-D_m(\wh\om)\la_{a\hat 3}$. For the modified susy transformation including $\la_{a\hat 3}=-\epbar_{+}\psi_{a-}$, the supercovariant extrinsic curvature is therefore
\bea
\label{scKma}
\wh K_{m a} \equiv \wh\om_{m a\hat 3}-\half\psibar_{m+}\psi_{a-}
=K_{m a}+\qter(\psibar_m\ga_a\psi_{\hat 3}+\psibar_a\ga_m\psi_{\hat 3}-\psibar_m\psi_a)
\eea
and as the bosonic part is symmetric,\footnote{
Use (\ref{CMNA}) and $C_{a b\hat 3}=0$, which is the case in the gauge $e_m{}^{\hat 3}=e_a{}^3=0$.}
$K_{a b}=K_{b a}$, we find that $\wh K_{a b}=\wh K_{b a}$. Performing the following decomposition,
\bea
\label{kappapm}
\ba[b]{rclcrcl}
\wh\om_{m a b} &=& \wh\om_{m a b}^{+}+\kappa_{m a b}^{-}, &&
\kappa_{m a b}^{-} &=& \dst\qter(\psibar_{m-}\ga_a\psi_{b-}
-\psibar_{m-}\ga_b\psi_{a-}+\psibar_{a-}\ga_m\psi_{b-}) \\[8pt]
\wh\om_{m a b}^{+} &=& \om(e)_{m a b}+\kappa_{m a b}^{+}, &&
\kappa_{m a b}^{+} &=& \dst\qter(\psibar_{m+}\ga_a\psi_{b+}
-\psibar_{m+}\ga_b\psi_{a+}+\psibar_{a+}\ga_m\psi_{b+})
\ea
\eea
we observe that $\wh\om_{m a b}^{+}$ is the supercovariant spin connection for the induced vielbein $e_m{}^a$. (Supercovariant both under ordinary susy transformations and under modified susy transformations.)

Fierzing in $d=4$ is done using the following formula
\bea
(\etabar\la)(\epbar\psi)=-\frac{1}{4}(\etabar O_j \psi)(\epbar O^j \la), \quad
O_j=(1, \;\; \ga_A, \;\; \frac{i}{\sqrt 2}\ga_{A B},  \;\; i\ga_5\ga_A,  \;\; \ga_5)
\eea
where in $O^j$ the Lorentz indices $A,B$ are raised. In addition one uses that, for Majorana spinors, $\epbar\ga^{A_1}\dots\ga^{A_k}\psi=(-)^k\psibar\ga^{A_k}\dots\ga^{A_1}\ep$ and $\epbar\ga_5\ga^{A_1}\dots\ga^{A_k}\psi=\psibar\ga_5\ga^{A_k}\dots\ga^{A_1}\ep$. With the decomposition $A=(a,\hat 3)$, we can write
\bea
O_j=(1, \;\; \ga_a, \;\; \ga_{\hat 3}, \;\; \frac{i}{\sqrt 2}\ga_{a b},  \;\; 
i\ga_{a\hat 3},  \;\; i\ga_5\ga_a,  \;\; i\ga_5\ga_{\hat 3},  \;\; \ga_5)
\eea
Using $\ga^{a b}=-\ep^{a b c}i\ga_5\ga_c\ga_{\hat 3}$, $\ep^{a b c}\equiv\ep^{a b c\hat 3}$ and $\ep_{a b k}\ep^{a b c}=-2\da_k{}^c$ we find that 
\bea
O_j=O_j^{+} \oplus O_j^{-}, \quad
O_j^{+} &=& (\ga_a, \;\; i\ga_a\ga_{\hat 3}; \;\; \ga_5, \;\; i\ga_5\ga_{\hat 3}) \nn\\
O_j^{-} &=& (1, \;\; \ga_{\hat 3}; \;\; i\ga_5\ga_a, \;\; i\ga_5\ga_a\ga_{\hat 3})
\eea
so that only $\etabar_{+}O_j^{+}\psi_{+}$, $\etabar_{-}O_j^{+}\psi_{-}$ and $\etabar_{+}O_j^{-}\psi_{-}$, $\etabar_{-}O_j^{-}\psi_{+}$ are nonvanishing.

A tensor with complete antisymmetrization in $d=3$ indices $a,b,c$ must be proportional to the $d=3$ Levi-Civita tensor $\ep_{a b c}$. For a tensor satisfying $C_{a b c}=-C_{c b a}$ this means
\bea
C_{[a b c]}=\frac{1}{3}(C_{a b c}+C_{b c a}+C_{c a b})=\ep_{a b c}C, \quad
C\equiv -\frac{1}{6}\ep^{a b c}C_{a b c}
\eea
Applying this to $C_{a b c}=\psibar_{a}\ga_b\psi_{c}$ and using $\ep^{a b c}\ga_c=i\ga_5\ga^{a b}\ga_{\hat 3}$, we find the following identity
\bea
\label{pgpident}
\psibar_a\ga_b\psi_c+\psibar_b\ga_c\psi_a+\psibar_c\ga_a\psi_b
=3 \ep_{a b c} C, \quad
C=\frac{1}{6} \psibar_a i\ga_5\ga^{a b}\ga_{\hat 3}\psi_b
\eea

With these conventions and tricks, let us now perform some of the technical derivations referred to in the main text. To prove (\ref{final-ep3}), we need to work out
\bea
\wt\ep=\half\ga^{a\hat 3}\ep_{1+}(-\epbar_{2+}\psi_{a-})
+\frac{3}{4}i\ga_5\ep_{1+}\Big( -\frac{2}{3}\epbar_{2+}i\ga_5\ga^a\psi_{a-} \Big)
-(1\lra 2)
\eea
Forming a scalar by multiplying with a spinor $\phibar$ and using $\ga_{a\hat 3}\ep_{+}=\ga_a\ga_{\hat 3}\ep_{+}=\ga_a\ep_{+}$, we get
\bea
\phibar\wt\ep=-\half(\phibar\ga^a\ep_{1+})(\epbar_{2+}\psi_{a-})
+\half(\phibar\ga_5\ep_{1+})(\epbar_{2+}\ga_5\ga^a\psi_{a-})
-(1\lra 2)
\eea
Fierzing this expression into the form $\epbar_{2+}O_j\ep_{1+}$, we find
\bea
\phibar\wt\ep=\frac{1}{8}(\epbar_{2+}O_j\ep_{1+})\Big[
\phibar\ga^a O_j\psi_{a-}-\phibar\ga_5 O_j\ga_5\ga^a\psi_{a-} \Big]
-(1\lra 2)
\eea
Only $O_j^{+}$ survives in $\epbar_{2+}O_j\ep_{1+}$; $\ga_5$ dependent terms in $O_j^{+}$ drop out due to ``$1\lra 2$.'' The remaining two objects in $O_j^{+}$, $\ga^a$ and $i\ga^a\ga_{\hat 3}$, contribute equally, and yield
\bea
\phibar\wt\ep=\frac{1}{4}(\epbar_{2+}\ga_b\ep_{1+})\Big[
\phibar\ga^a\ga^b\psi_{a-}-\phibar\ga_5\ga^b\ga_5\ga^a\psi_{a-} \Big]
-(1\lra 2)
\eea
Using $\ga_5\ga^b\ga_5=-\ga^b$, $\ga^a\ga^b+\ga^b\ga^a=2\eta^{a b}$ and $\epbar_{1+}\ga_b\ep_{2+}=-\epbar_{2+}\ga_b\ep_{1+}$, we find
\bea
\wt\ep=(\epbar_{2+}\ga^a\ep_{1+})\psi_{a-}=\half\xi^a\psi_{a-}
\eea
where $\xi^a=2(\epbar_{2+}\ga^a\ep_{1+})$. This proves (\ref{final-ep3}).

To prove (\ref{la3ab}), we first find, using same tricks while Fierzing, that
\bea
\la_{a\hat 3}(\ep_{2+})\la_{b\hat 3}(\ep_{1+})-(1\lra 2)
&=& (\epbar_{2+}\psi_{a-})(\epbar_{1+}\psi_{b-})-(1\lra 2) \nn\\
&=& -\qter(\epbar_{2+}O_j\ep_{1+})(\psibar_{b-}O^j\psi_{a-})-(1\lra 2) \nn\\
&=& -(\epbar_{2+}\ga^c\ep_{1+})(\psibar_{b-}\ga_c\psi_{a-})
=\half\xi^c(\psibar_{a-}\ga_c\psi_{b-})
\eea
Writing $\wh\om_{m a b}=\wh\om_{m a b}^{+}+\kappa_{m a b}^{-}$ as in (\ref{kappapm}), we find from (\ref{composites}) that
\bea
(\la_3)_{a b}=\xi^c\wh\om_{c a b}^{+}-\epbar_{[2+}\ga_{a b}\zeta_{-}(\ep_{1]+})
+\xi^c\kappa_{c a b}^{-}-\half\xi^c(\psibar_{a-}\ga_c\psi_{b-})
\eea
Using the identity in (\ref{pgpident}) with $\ga_{\hat 3}\psi_{b-}=-\psi_{b-}$, we obtain
\bea
\label{epsW}
\kappa_{c a b}-\half\psibar_{a-}\ga_c\psi_{b-}
=\qter(\psibar_{c-}\ga_a\psi_{b-}-\psibar_{c-}\ga_b\psi_{a-}-\psibar_{a-}\ga_c\psi_{b-})
=-\half\ep_{a b c} W
\eea
where $W\equiv \qter\psibar_{a-}i\ga_5\ga^{a b}\psi_{b-}$. Using $\ep_{a b c}\ga^c=i\ga_5\ga_{a b}\ga_{\hat 3}$, we find
\bea
\xi^c\ep_{a b c}=2(\epbar_{2+}i\ga_5\ga_{a b}\ep_{1+})
=\epbar_{[2+}\ga_{a b}i\ga_5\ep_{1]+}
\eea
This allows to write $(\la_3)_{a b}$ in the following form
\bea
(\la_3)_{a b}=\xi^c\wh\om_{c a b}^{+}-\epbar_{[2+}\ga_{a b}\zeta^\rr_{-}(\ep_{1]+})
\eea
where $\zeta^\rr_{-}(\ep_{+})=\zeta_{-}(\ep_{+})+\half i\ga_5\ep_{+}W$. This proves (\ref{la3ab}).

To prove (\ref{dapsiam}), we first collect the terms in (\ref{dapsiam1}) remaining after projection with $P_{-}$,
\bea
\da^\rr(\ep_{+})\psi_{a-} &=& -(\epbar_{+}\ga^b\psi_{a+})\psi_{b-}
+\ga^{c\hat 3}\ep_{+}\wh\om_{a c\hat 3}
-\frac{3}{2}i\ga_5\ep_{+}A_a
-\ga_a\zeta_{+}(\ep_{+}) \nn\\
&& +\half\ga_{c\hat 3}\psi_{a+}(-\epbar_{+}\psi^c_{-})
+\frac{3}{4}i\ga_5\psi_{a+}\Big(-\frac{2}{3}\epbar_{+}i\ga_5\ga^c\psi_{c-} \Big)
\eea
where we used, in particular, that $P_{-}\ga_5=\ga_5 P_{+}$ as $\ga^{\hat 3}\ga_5=-\ga_5\ga^{\hat 3}$. Completing $\wh\om_{a c\hat 3}$ into the $\da^\rr(\ep_{+})$ supercovariant $\wh K_{a c}=\wh\om_{a c \hat 3}-\half\psibar_{a+}\psi_{c-}$, see (\ref{scKma}), we write
\bea
\da^\rr(\ep_{+})\psi_{a-} &=& \ga^b\ep_{+}\wh K_{a b}
-\frac{3}{2}i\ga_5\ep_{+} A_a-\ga_a\zeta_{+}(\ep_{+})+Q_{-} \nn\\
Q_{-} &\equiv& \half\ga^b\ep_{+}(\psibar_{a+}\psi_{b-})
-\psi_{b-}(\epbar_{+}\ga^b\psi_{a+}) \nn\\
&& -\half\ga^b\psi_{a+}(\epbar_{+}\psi_{b-})
+\half\ga_5\psi_{a+}(\epbar_{+}\ga_5\ga^b\psi_{b-})
\eea
Fierzing $\phibar_{+} Q_{-}$ into the form $\psibar_{a+}O_j\psi_{b-}$, where only $O_j^{-}$ survives (with $\ga_{\hat 3}$ and $i\ga_5\ga_c\ga_{\hat 3}$ doubling the contributions of $1$ and $i\ga_5\ga_c$, respectively), gives
\bea
\phibar_{+}Q_{-} &=& 
(\psibar_{a+}\psi_{b-})(\phibar_{+}\ga^b\ep_{+})\Big( \half-\half+\qter-\qter \Big) \nn\\
&& +(\psibar_{a+}\ga_5\ga_c\psi_{b-})\Big(
\half\phibar_{+}\ga_5\ga^c\ga^b\ep_{+}
+\qter\phibar_{a+}\ga_5\ga^b\ga^c\ep_{+}
-\qter\phibar_{+}\ga_5\ga^c\ga^b\ep_{+}
\Big) \nn\\
&=& (\psibar_{a+}\ga_5\ga_c\psi_{b-})\Big( 
\half\phibar_{+}\ga_5(\ga^c \ga^b-\ga^{c b})\ep_{+} \Big)
=\half(\phibar_{+}\ga_5\ep_{+})(\psibar_{a+}\ga_5\ga^b\psi_{b-}) \qquad
\eea
We can absorb $Q_{-}$ by redefining $A_a$,
\bea
\da^\rr(\ep_{+})\psi_{a-} &=& \ga^b\ep_{+}\wh K_{a b}
-\frac{3}{2}i\ga_5\ep_{+} \wh A_a-\ga_a\zeta_{+}(\ep_{+})
\eea
where $\wh A_a=A_a+\frac{1}{3}\psibar_{a+}i\ga_5\ga^b\psi_{b-}$. This proves (\ref{dapsiam}).

Finally, to prove (\ref{dapsiap}), we first collect the terms in (\ref{dapsiap1}) which survive the projection with $P_{+}$,
\bea
\da^\rr(\ep_{+})\psi_{m+} &=& 2 D^\rr(\wh\om^{+})_m\ep_{+}-\ga_m\zeta_{-}(\ep_{+})
+\half\ga^{a b}\ep_{+}\kappa_{m a b}^{-}+Q_{+} \nn\\
Q_{+} &\equiv& \half\ga^a\psi_{m-}(\epbar_{+}\psi_{a-})
+\half\ga_5\psi_{m-}(\epbar_{+}\ga_5\ga^a\psi_{a-})
\eea
Fierzing $\phibar_{-}Q_{+}$ into the form $\psibar_{a-}O_j\psibar_{m-}$, where only $Q_j^{+}$ survives (with $i\ga_c\ga_{\hat 3}$ and $i\ga_5\ga_{\hat 3}$ doubling the contributions of $\ga_c$ and $\ga_5$, respectively), gives
\bea
\phibar_{-}Q_{+} &=& (\psibar_{a-}\ga_c\psi_{m-})\Big[
-\qter(\phibar_{-}\ga^a\ga^c\ep_{+})
+\qter(\phibar_{-}\ga^c\ga^a\ep_{+}) \Big] \nn\\
&& +(\psibar_{a-}\ga_5\psi_{m-})(\phibar_{-}\ga^a\ga_5\ep_{+})\Big( -\qter+\qter \Big)
\eea
so that $Q_{+}=-\half\ga^{a b}\ep_{+}(\psibar_{a-}\ga_b\psi_{m-})$. Combining with the $\kappa_{m a b}^{-}$ term, we find
\bea
\half\ga^{a b}\ep_{+}\kappa_{m a b}^{-}+Q_{+}
&=& -\frac{1}{8}\ga^{a b}\ep_{+}(\psibar_{m-}\ga_a\psi_{b-}-\psibar_{m-}\ga_b\psi_{a-}
-\psibar_{a-}\ga_m\psi_{b-}) \nn\\
&=& -\frac{1}{2}\ga^{a b}\ep_{+}\Big(-\half\ep_{a b m} W \Big)
=-\half \ga_m i\ga_5\ep_{+} W
\eea
with $W\equiv \qter\psibar_{a-}i\ga_5\ga^{a b}\psi_{b-}$. We used the result in (\ref{epsW}) and $\ga^{a b}\ep_{a b c}=2i\ga_5\ga_c\ga_{\hat 3}$. We can now combine this result with the term $-\ga_m\zeta_{-}(\ep_{+})$ to find
\bea
\da^\rr(\ep_{+})\psi_{m+} &=& 2 D^\rr(\wh\om^{+})_m\ep_{+}-\ga_m\zeta^\rr_{-}(\ep_{+})
\eea
where $\zeta^\rr_{-}(\ep_{+})=\zeta_{-}(\ep_{+})+\half i\ga_5\ep_{+} W$. This proves (\ref{dapsiap}).



\end{document}